\documentclass[aps, prd, 10pt, notitlepage, twocolumn, superscriptaddress,nofootinbib, floatfix, longbibliography]{revtex4-2}

\pdfoutput=1

\PassOptionsToPackage{dvipsnames}{xcolor}

\usepackage{graphicx}
\usepackage{amsmath}
\usepackage{amssymb}
\usepackage{amsfonts}
\usepackage{mathtools}
\usepackage{svg}

\usepackage[T1]{fontenc}
\usepackage[utf8]{inputenc}
\usepackage{microtype}

\usepackage[normalem]{ulem} %

\usepackage{anyfontsize}

\usepackage[linktocpage,breaklinks]{hyperref}
\usepackage[all]{hypcap}
\usepackage{xcolor}

\usepackage{cleveref}

\Crefname{equation}{Eq.}{Eqs.} 

\usepackage{tensor}
\usepackage{bm}

\usepackage{graphicx}
\graphicspath{{./figures/}}

\usepackage{tikz}
\ifdefined\myext
\usetikzlibrary{external}
\fi

\usepackage{natbib}
\usepackage{orcidlink}
\usepackage{verbatim}

\hypersetup{colorlinks=true,
            citecolor=RoyalBlue,
            linkcolor=RoyalBlue,
            urlcolor=RoyalBlue}

\usepackage{xspace}
\usepackage{dsfont}
\usepackage{soul} 
\usepackage{dsfont}
\usepackage{multirow}
\usepackage{cancel}

\usepackage[caption=false]{subfig}

\usepackage{stackengine}
\usepackage{scalerel}

\stackMath

\newcommand{\beq}{\begin{equation}}
\newcommand{\eeq}{\end{equation}}

\newcommand{\nrpyell}{{\texttt{NRPyElliptic}}\xspace}

\newcommand{\pinndorama}{{\texttt{Pinndorama}}\xspace}

\newcommand{\SinhSymTP}{{\texttt{SinhSymTP}}\xspace}
\newcommand{\nrpy}{{\texttt{NRPy}}\xspace}
\newcommand{\TwoPunctures}{{\texttt{TwoPunctures}}\xspace}

\newcommand{\xx}{\mathtt{x}}           
\newcommand{\xxzero}{\mathtt{x_{1}}}   
\newcommand{\xxone}{\mathtt{x_{2}}}    
\newcommand{\xxtwo}{\mathtt{x_{3}}}    
\newcommand{\Nzero}{\mathtt{N_{1}}}    
\newcommand{\None}{\mathtt{N_{2}}}     
\newcommand{\Ntwo}{\mathtt{N_{3}}}     
\renewcommand{\eqref}[1]{Eq.\,(\ref{#1})}
\newcommand{\eqsref}[1]{Eqs.\,(\ref{#1})}

\DeclareMathOperator*{\argmin}{argmin}

\definecolor{amethyst}{rgb}{0.6, 0.4, 0.8}

\begin{document}
\date{October 20, 2026}
\title{Binary Black Hole Puncture Initial Data with Physics-Informed Neural Networks}
\author{Thiago Assump\c c\~ao\,\orcidlink{0000-0002-3419-892X}}
\email{thiago.assumpcao@nottingham.ac.uk}
\affiliation{Center for Gravitation, Cosmology and Astrophysics, Department of Physics \& Astronomy, University of Wisconsin-Milwaukee, Milwaukee, WI 53211, USA}
\affiliation{Nottingham Centre of Gravity, Nottingham NG7 2RD, UK}
\affiliation{School of Mathematical Sciences, University of Nottingham, Nottingham, NG7 2RD, UK}

\author{Ignacio Brevis\,\orcidlink{0000-0003-1620-019X}}
\email{ignacio.brevis1@nottingham.ac.uk}
\affiliation{School of Mathematical Sciences, University of Nottingham, Nottingham, NG7 2RD, UK}

\author{Irene Pitsiladi\,\orcidlink{0009-0009-3022-0614}}
\email{i.b.pitsiladi@qmul.ac.uk}
\affiliation{School of Mathematical Sciences, University of Nottingham, Nottingham, NG7 2RD, UK}
\affiliation{School of Mathematical Sciences, Queen Mary University of London, London, E1 4NS, UK}

\author{Martin Zinzen\,\orcidlink{0009-0001-7690-0326}}
\email{mzinzen1@sheffield.ac.uk}
\affiliation{School of Mathematical Sciences, University of Nottingham, Nottingham, NG7 2RD, UK}
\affiliation{School of Mathematical and Physical Sciences, University of Sheffield, Sheffield, S10 2TN, UK}

\author{Miguel Bezares\,\orcidlink{0000-0002-3739-6980}}
\email{miguel.bezaresfigueroa@nottingham.ac.uk}
\affiliation{Nottingham Centre of Gravity, Nottingham NG7 2RD, UK}
\affiliation{School of Mathematical Sciences, University of Nottingham, Nottingham, NG7 2RD, UK}

\begin{abstract}
We present \pinndorama, a physics-informed neural network (PINN) code for solving the Hamiltonian constraint in the puncture formalism. The networks are trained solely on the governing equation and boundary conditions, without an analytical or perturbative guiding solution, with collocation points sampled in curvilinear coordinates. Training combines two optimization algorithms: an initial stage using adaptive moment estimation (Adam), followed by the self-scaled Broyden quasi-Newton method. We apply this approach to a spherically symmetric test problem with a known exact solution, axisymmetric binary punctures with unequal spins, and a fully three-dimensional GW150914-like configuration with unequal spins and orbital momenta. Comparisons with the \nrpyell solver yield relative $L^2$ errors of order $10^{-8}$ in axisymmetry and $10^{-5}$ in three dimensions.
In addition, we reduce the hyperbolic-relaxation time in \nrpyell by a factor of approximately $4.6$ by training a parametric network that takes the puncture spin as input and provides a physics-informed initial guess to the solver. These results demonstrate that PINNs can provide accurate binary-puncture initial data and complement conventional elliptic solvers in numerical relativity.
\end{abstract}

\maketitle

\section{Introduction}

The direct detection of gravitational waves (GWs) from compact-binary mergers has become a powerful channel to test general relativity, probe fundamental physics in strong gravity, and investigate matter at high densities~\cite{testsGR_GWTC-1, testsGR_GWTC-2, testsGR_GWTC-3, NSNS_2019}. The LIGO--Virgo--KAGRA collaboration has reported nearly 400 compact-binary coalescence candidates~\cite{GWTC-5.0}, predominantly associated with binary black hole (BBH) mergers. These BBH systems are expected to remain important sources for current and future ground- and space-based GW detectors, including the Einstein Telescope~\cite{ET:2025xjr}, Cosmic Explorer~\cite{cosmic_explorer_us_contribution}, and LISA~\cite{LISA3}.

Extracting physical parameters from GW observations requires accurate waveform models, many of which are informed by numerical relativity (NR) simulations~\cite{SXS_collab}. These simulations begin with initial data describing the compact binary system. Constructing such data requires solving the Hamiltonian and momentum constraints of general relativity~\cite{Tichy_2016}, which can be formulated as a system of nonlinear elliptic partial differential equations (PDEs). The resulting constraint-satisfying data provide the starting point for numerical evolutions of Einstein's equations, from which the gravitational-wave signal is extracted. Efficient and accurate methods for solving the constraints are therefore an essential component of NR simulations.

Several mathematical formulations and numerical techniques have been developed to solve Einstein's equations~\cite{numerical-methods}. Widely used initial data and evolution solvers predominantly employ finite-difference or spectral methods; see Refs.~\cite{LISAConsortiumWaveformWorkingGroup:2023arg,Hilditch:2024nhf} for an overview of available codes.  Recent advances in artificial intelligence, supported by improvements in hardware and algorithms, motivate investigating whether these approaches can complement established numerical methods in NR.

Physics-Informed Neural Networks (PINNs) are a class of machine-learning methods that incorporate the physical laws of a problem into neural network training. PINNs have been employed in a wide range of fields~\cite{PINN_usecases_review}. These include areas such as geophysics~\cite{nonlinear_structures_PINN, hydrology_PINN}, hydrodynamics~\cite{espresso_cup_PINN, turbulence_PINN, turbulence_PINN2, fluid_stress_PINN}, electrostatics~\cite{molecular_electrostatics_PINN, charged_particles_PINN}, materials science~\cite{cracks_PINN, material_properties_PINN}, and biomedicine~\cite{cell_signalling_PINN, systems_biology_PINN, cardiac_PINN, SVIHR_PINN}.

A neural network represents a function whose parameters, referred to as weights and biases, are adjusted through training. The aim is to minimize a loss function that quantifies how closely the network output matches the desired behavior~\cite{deep_learning_intro}. At each optimization step, the network is evaluated on a set of inputs and the loss is computed from its outputs. Gradients of the loss with respect to the network parameters are obtained through backpropagation and used by an optimization algorithm to update those parameters.

In PINNs, the governing differential equations and boundary conditions are incorporated into the loss function, which may also include contributions from numerical data~\cite{pinns-intro}. Training seeks a network that approximates the physical solution by minimizing violations of these equations and conditions. This framework allows training without requiring reference solutions, making it particularly relevant to the constraint equations of general relativity, where initial data must often be constructed for configurations whose solutions are not known in advance.

Neural networks have been applied to several areas of computational physics, including the solution to the Riemann problem~\cite{Riemann_problem_PINN}, magnetohydrodynamic (MHD) simulations~\cite{MHD_PINN, MHD_NO, low-dimensional-models-MHD}, and conservative-to-primitive recovery in general-relativistic MHD simulations~\cite{ML_GRMHD}. In gravitational physics, applications include black hole perturbation equations~\cite{pinns-bh-perturbation-eqns, pinns-schwarzschild-perturbation, Luna:2022rql,Cornell:2024azz}, quasinormal modes in modified gravity~\cite{Luna:2024spo}, and relativistic corrections to Newtonian and post-Newtonian approximations inferred from GW data~\cite{Newtonian_correction_PINN_from_data} and NR waveforms~\cite{PN_PINN_from_NR}, respectively. PINNs have also been used to study gravitational collapse in spherical symmetry~\cite{Ferrer-Sanchez:2025cns}. More recently, Carretero et al.~\cite{carretero2026solvinggeneralrelativityphysicsinformed} applied PINNs to standard NR benchmarks and the three-dimensional evolution of a solitonic boson star, providing the first PINN solution of the coupled Einstein--Klein--Gordon equations. Shukla et al.~\cite{shukla2026punctureevolutionphysicsinformedneural} studied the moving-puncture evolution of a single nonspinning black hole in spherical symmetry, using early-time and outer-boundary numerical data together with the vacuum Einstein equations and gauge conditions.

For initial data construction, Zhou et al.~\cite{619s-p6md} solved the Hamiltonian constraint for binary black holes by incorporating an approximate analytical solution into the network ansatz and training with the Adam optimizer. They used uniform random sampling within a spherical domain of radius $R_{\max}=30$, with $20000$ interior points and $8000$ boundary points. For their unequal-mass spinning configuration, they report a mean relative $L^2$ error of approximately $9\times10^{-2}$ over five independent runs.

In this paper, we present \pinndorama, a physics-informed neural network code for constructing binary black hole puncture initial data in full three dimensions, whose source code is publicly available\footnote{\url{https://github.com/assumpcaothiago/Pinndorama}}. We adopt the conformal transverse-traceless (CTT) decomposition of the constraint equations with a conformally flat background~\cite{York73} and learn the regular puncture correction without embedding an approximate analytical guiding solution in the network ansatz. Our networks are trained solely on the governing equation and boundary conditions, without using reference solutions as training data. We use cell-centered sampling in puncture-adapted \SinhSymTP coordinates and Adam followed by self-scaled Broyden optimization. We assess the method through a one-dimensional problem with a known exact solution, an axisymmetric binary-puncture configuration, and a fully three-dimensional configuration with unequal spins and orbital momenta. The one-dimensional solution is compared with its analytical reference, while the puncture solutions are validated against \nrpyell, which has itself been validated against \texttt{TwoPunctures}~\cite{assumpcao_fast_2022}. We obtain relative $L^2$ errors of order $10^{-8}$ in axisymmetry and $10^{-5}$ for the selected three-dimensional calculation. We also develop a parametric axisymmetric network that takes the common spin of the two punctures as an additional input. Using this network as an initial guess for \nrpyell reduces the subsequent hyperbolic relaxation time by a factor of approximately $4.6$.

The remainder of the paper is organized as follows. In section ~\ref{sec:punct}, we summarize the CTT formulation. In section~\ref{sec:pinns}, we provide a brief and self-contained introduction to neural networks, physics-informed neural networks, and a discussion on the convergence of physics-informed neural networks. Section~\ref{sec:code_implementation} describes the code implementation, the puncture-adapted coordinate
systems used to sample the collocation points, the boundary conditions, the loss function, and the
training strategy. The results and validation against reference solutions are presented in section~\ref{sec:results}. In section~\ref{sec:conclusions}, we provide our conclusions and plans for future work.

\section{The puncture formalism}\label{sec:punct}

We adopt the $3+1$ decomposition of spacetime~\cite{Gourgoulhon:2007ue}, in which the line element is
\begin{equation}
    ds^2 = -\alpha^2 dt^2 + \gamma_{ij}(dx^i + \beta^i dt)(dx^j + \beta^j dt)\,,
\end{equation}
where $\alpha$ is the lapse function, $\beta^i$ is the shift vector, and $\gamma_{ij}$ is the metric induced on each spatial hypersurface $\Sigma_t$. In vacuum, the initial data $(\gamma_{ij},K_{ij})$ must satisfy the Hamiltonian and momentum constraints,
\begin{align}
    R + K^2 - K_{ij}K^{ij} &= 0 \,, \label{eqn:hamiltonian_constraint}\\
    \nabla_j(K^{ij}-\gamma^{ij}K) &= 0 \,, \label{eqn:momentum_constraint}
\end{align}
where $R$ is the spatial Ricci scalar, $K_{ij}$ is the extrinsic curvature, $K=\gamma^{ij}K_{ij}$ is its trace, and $\nabla_i$ is the covariant derivative compatible with $\gamma_{ij}$.

The conformal transverse-traceless (CTT) decomposition~\cite{Cook:2000vr} introduces the conformal factor $\psi$ and conformal metric $\tilde{\gamma}_{ij}$ through
\begin{equation}
    \gamma_{ij}=\psi^4\tilde{\gamma}_{ij}\,,
    \qquad
    K_{ij}=\psi^{-2}\tilde{A}_{ij}+\frac13\gamma_{ij}K\,.
\end{equation}
Here, $\tilde{A}_{ij}$ is trace-free with respect to $\tilde{\gamma}_{ij}$, which also raises and lowers conformal indices. We write
\begin{equation}
    \tilde{A}^{ij}=(\tilde{\mathds{L}}V)^{ij}+\tilde{M}^{ij}\,,
\end{equation}
where $\tilde{M}^{ij}$ is a freely specified symmetric trace-free tensor and the longitudinal operator is
\begin{equation}
    (\tilde{\mathds{L}}V)^{ij}
    =\tilde{\nabla}^iV^j+\tilde{\nabla}^jV^i
    -\frac23\tilde{\gamma}^{ij}\tilde{\nabla}_kV^k\,.
\end{equation}
The derivative $\tilde{\nabla}_i$ is compatible with $\tilde{\gamma}_{ij}$. The constraints then become
\begin{align}
    \tilde{\nabla}^2\psi-\frac18\tilde{R}\psi
    -\frac1{12}K^2\psi^5
    +\frac18\tilde{A}_{ij}\tilde{A}^{ij}\psi^{-7} &= 0 \,,\\
    \tilde{\nabla}_j(\tilde{\mathds{L}}V)^{ij}
    -\frac23\psi^6\tilde{\nabla}^iK
    +\tilde{\nabla}_j\tilde{M}^{ij} &= 0 \,,
\end{align}
where $\tilde{\nabla}^2=\tilde{\nabla}_i\tilde{\nabla}^i$ and $\tilde{R}$ is the conformal Ricci scalar.

For the Bowen--York puncture data considered here, we choose a maximal initial hypersurface, $K=0$, conformal flatness, $\tilde{\gamma}_{ij}=\hat{\gamma}_{ij}$, and $\tilde{M}^{ij}=0$, together with asymptotic flatness, $\psi\to1$ at spatial infinity. Here, $\hat{\gamma}_{ij}$ is the flat spatial metric, whose Ricci tensor vanishes. The constraints reduce to
\begin{align}
    \hat{\nabla}^2\psi+\frac18\psi^{-7}\tilde{A}_{ij}\tilde{A}^{ij} &= 0 \,,\\
    \hat{\nabla}^2V^i+\frac13\hat{\nabla}^i(\hat{\nabla}_jV^j) &= 0 \,,
\end{align}
with $\hat{\nabla}_i$ the covariant derivative compatible with $\hat{\gamma}_{ij}$.

The momentum constraint admits the analytical Bowen--York solution~\cite{Bowen:1980yu}. For $N_p$ punctures, the vector potential is
\begin{equation} \label{eq:general-bowen-york-solution}
        \vec{V} = \sum_{n=1}^{N_p} \left( - \frac{7}{4 |\vec{x}_n|} \vec{P}_n - \frac{\vec{x}_n {\cdot} \vec{P}_n}{4 |\vec{x}_n|^3} \vec{x}_n + \frac{1}{|\vec{x}_n|^3} \vec{x}_n {\times} \vec{S}_n \right)\,,
\end{equation}
where $\vec{x}_n=\vec{x}-\vec{x}_n^{\,(\mathrm{p})}$ is the displacement from puncture $n$, located at $\vec{x}_n^{\,(\mathrm{p})}$, and $\vec{P}_n$ and $\vec{S}_n$ specify its linear momentum and spin.

The remaining Hamiltonian constraint must be solved numerically. To isolate the singular behavior of $\psi$ at the punctures, we follow Ref.~\cite{Brandt:1997tf} and write
\begin{equation}\label{eqn:non-singular_puncture_constraint}
    \psi=\psi_{\rm singular}+u\,,
    \qquad
    \psi_{\rm singular}=1+\sum_{n=1}^{N_p}\frac{m_n}{2|\vec{x}_n|}\,,
\end{equation}
where $m_n$ is the bare-mass parameter of puncture $n$. We require $u$ to remain regular at the punctures and to approach zero as $|\vec{x}|\to\infty$. Since $\hat{\nabla}^2\psi_{\rm singular}=0$ away from the punctures, the Hamiltonian constraint becomes
\begin{equation}\label{eq:general-Hamiltonian}
    \hat{\nabla}^2u+\frac18\tilde{A}_{ij}\tilde{A}^{ij}
    (\psi_{\rm singular}+u)^{-7}=0\,.
\end{equation}
This nonlinear elliptic equation is the binary-puncture problem solved by our PINNs.

We also consider a manufactured, spherically symmetric problem that retains the nonlinear structure of~\eqref{eq:general-Hamiltonian} while admitting an exact solution. Its governing equation is
\begin{equation}\label{eq:1d-toy-pde}
    \partial_r^2u+\frac2r\partial_ru
    +\frac{A(r)}{8[\psi_{\rm singular}(r)+u]^7}=0\,,
\end{equation}
where $\psi_{\rm singular}(r)=1+m/(2r)$. Setting $m=1$, we construct $A(r)$ by substituting
\begin{equation}\label{eq:1d_exact_solution}
    u_{\rm ex}(r)=\frac{2+2r}{r^2+2r+2}
\end{equation}
into~\eqref{eq:1d-toy-pde}. This solution satisfies regularity at the origin, $\partial_ru(0)=0$, and $u\to0$ as $r\to\infty$. This one-dimensional problem provides a controlled test of the numerical method before its application to binary puncture initial data.

\section{Physics-informed neural networks}\label{sec:pinns}

This section provides a self-contained introduction to the mathematical framework underlying physics-informed neural networks~\cite{pinns-intro}.
We begin by reviewing the architecture and training of fully connected neural networks in Sec.~\ref{sec:nn_formalism}, before describing how physical laws are incorporated into the learning process through the loss function in the PINN setting in Sec.~\ref{sec:pinn_formalism}. We close with a discussion of the three sources of error that govern convergence of the method to the true PDE solution in Sec.~\ref{sec:pinns_convergence}.

\subsection{Neural network formalism}
\label{sec:nn_formalism}

A fully connected feed-forward neural network $nn_\theta:\mathbb{R}^{n_0}\to \mathbb{R}^{n_{M+1}}$ with $M$ hidden layers is defined by:
\begin{equation}
nn_\theta(x)\coloneq L_{M+1}\circ L_{M}\circ L_{M-1}\circ \cdots \circ L_1(x).
\end{equation}
Each layer $L_i$ 
is given by the composition of an activation function, $\sigma$ or $\tilde \sigma$, with an affine function $\ell_i$
    \begin{equation}
    L_{i}(x) =
        \begin{cases}
        \sigma \circ \ell_i(x) = \sigma(W_i x + b_i), & 1 \leq i \leq M\,, \\
        \tilde\sigma \circ \ell_i(x) = \tilde\sigma(W_i x + b_i), & i = M+1\,,
        \end{cases}
    \end{equation}
where $W_i$ is an ${n_{i}\times n_{i-1}}$ matrix and $b_i\in \mathbb{R}^{n_i}$. The first $M$ layers are known as the hidden layers of the neural network, $\sigma$ denotes a nonlinear function, and $\tilde\sigma$ is an activation function whose nature varies with the application (it may be linear or even the identity function). The nonlinearity of the activation functions in the hidden layers is essential for ensuring good approximability properties in the neural network (also known as expressivity). For a more detailed description on this subject, see, e.g., Ref.~\cite{Pinkus_1999}.

The collection of all elements in the matrices and vectors $\{W_1,b_1,W_2,b_2,\dots,W_{M+1}, b_{M+1}\}$ is commonly denoted by $\theta$, representing the trainable parameters in the neural network. The training of a neural network is given by the minimization of a functional or loss function $\mathcal{L}$ that quantifies the discrepancy between the neural network and the desired behavior. Then, the trained parameters, denoted by $\theta^*$, will be obtained by
\begin{equation}
\theta^* = \argmin_{\theta}\mathcal{L}(\theta) \,.
\end{equation}
Depending on the application and the data available, we can define different loss functions. Some of the most common are mean squared error for regression problems, cross-entropy for classification, and Kullback-Leibler divergence for probabilistic applications (see, e.g., Ref.~\cite{Goodfellow-DL-2016}).

The loss function is, in most deep learning cases, non-linear and non-convex with respect to $\theta$. The minimization is normally performed iteratively via gradient-based methods, where the gradient $\nabla_\theta \mathcal{L}(\theta)$ is computed exactly (to machine precision) by backpropagation~\cite{Rumelhart1986}, a reverse-mode automatic differentiation algorithm~\cite{Baydin2017AD}.

\subsection{Physics-informed method}\label{sec:pinn_formalism}

When using neural networks for solving PDEs in a physics-informed neural networks~\cite{RaiPerKarJCP2019} setting, we approximate the PDE solution with a function $u_\theta$ constructed from a neural network $nn_\theta$, where the activation function in the final layer $\tilde\sigma$ is the identity.

In this case, the loss function is defined as the residual of the PDE in a least-squares sense. Let $u$ be the solution of the following problem with Dirichlet boundary condition:
\begin{equation}
    \left\{
    \begin{array}{r@{}ll}
        -\nabla^2 u + F(u) &{}= 0, &  \text{ in } \Omega\,, \\
        u &{}= 0, &  \text{ on } \partial\Omega\,,
    \end{array}
    \right.
    \label{eq:pde_elliptic_setting}
\end{equation}
where $\Omega$ is a bounded subset of $\mathbb{R}^n$ and $F$ is a nonlinear operator over $u$. The goal in the PINNs approach is that $u_\theta$ learns the \textit{physics} behind the PDE through the residual of the equation in the $L^2$ norm~\cite{sirignano_dgm_2018}:
\begin{equation}
    \mathcal{R}[u] = -\nabla^2 u + F(u)\,.
\end{equation}
The network $u_\theta$ is trained by minimizing a discrete approximation to $\|\mathcal{R}[u_\theta]\|^{2}_{L^{2}(\Omega)}$, commonly evaluated using Monte Carlo sampling:
\begin{equation}
    \mathcal{L}_{\text{pde}}(\theta) = \frac{|\Omega|}{N}\sum_{i=1}^{N} \left(-\nabla^2 u_{\theta}(x_i) + F(u_\theta(x_i))\right)^2,
\end{equation}
where the collocation points $\{x_i\}_{i=1}^{N}$ are randomly sampled from a uniform distribution in $\Omega$.
Although we focus on 
a uniform sampling strategy, many other alternatives are available in the literature, such as quadrature rules~\cite{RIVERA2022114710}, stochastic quadrature rules~\cite{TAYLOR20261}, importance sampling~\cite{Gao2023Failure}, and adaptive weights on the collocation points~\cite{MCCLENNY2023self}.

For a complete description of $u_\theta$ as an approximate solution of~\eqref{eq:pde_elliptic_setting}, the neural network expression must also learn the boundary conditions.
This can be achieved in two ways:
\begin{itemize}
    \item {\bf Strong imposition}: The boundary conditions are imposed on the neural-network-based approximation function. We define $u_\theta(x) \coloneq nn_\theta(x) b(x)$, where $b$ is called bubble function, which is usually a $C^{1}(\overline{\Omega})$ function, such that $b(x)\neq 0$ in the interior of $\Omega$ and $b|_{\partial\Omega} \coloneq 0$. In this setting, the full loss function is defined as
    $\mathcal{L}\coloneq \mathcal{L}_{\text{pde}}$.

    This idea can also be generalized to non-zero Dirichlet boundary conditions by including a lifting function into $u_\theta$ (see, e.g., Section 6.1 of Ref.~\cite{evans2010partial}).
    \item {\bf Weak imposition}: The boundary conditions are imposed as penalty terms. We define the following residual expression~\cite{sirignano_dgm_2018}
    \begin{equation}
        \mathcal{R}_{\text{bc}}[u] = u - g \,,
    \end{equation}
    where $g$ represents the boundary condition (in the above~\eqref{eq:pde_elliptic_setting}, $g=0$).
    The extra loss term is defined by a discrete (Monte Carlo) version of $\|\mathcal{R}_{\text{bc}}[u_\theta]\|^{2}_{L^{2}(\partial \Omega)}$:
    \begin{equation}
        \mathcal{L}_{\text{bc}}(\theta) \coloneq \frac{|\partial \Omega|}{N_{\text{bc}}}\sum_{i=1}^{N_{\text{bc}}} \left(u_\theta(x_i) - g(x_i)\right)^2,
    \end{equation}
    where $x_i$ is randomly sampled from a uniform distribution on $\partial\Omega$. In this case, $u_\theta$ is given directly by the neural network function, i.e., $u_\theta(x) \coloneq nn_\theta(x)$.

    When using weak imposition, the final loss is given by $\mathcal{L} = \mathcal{L}_{\text{pde}} + \lambda \mathcal{L}_{\text{bc}}$, with $\lambda$ a penalization parameter.
    In many practical applications, the constants given by the measure of $|\Omega|$ and $|\partial \Omega|$ are absorbed into $\lambda$, since their effect is a rescaling between the loss terms.
\end{itemize}

Strong imposition allows the training to focus only on the main term of the PDE, without adding extra terms into the loss. One caveat is that the strong imposition can be implemented only for Dirichlet boundary conditions.
In contrast, the weak imposition can also handle more general cases, such as Dirichlet, Neumann, and Robin conditions.
In this work, we consider the weak imposition case.

Beyond imposing boundary conditions, we can also incorporate prior knowledge about the qualitative or asymptotic behavior of the solution directly into the neural-network approximation~\cite{yoon2024,smithEikoNet}. In particular, we may consider an ansatz of the form
\begin{equation}
    u_\theta(x) := \zeta(x)nn_{\theta}(x) \,,
    \label{eqn:bubble_ansatz}
\end{equation}
where $\zeta$ is a non-trainable function chosen to encode known features of the solution.
Unlike the bubble functions introduced above to strongly impose Dirichlet data, $\zeta$ does not satisfy a boundary condition; instead, it embeds useful prior knowledge and simplifies the function the neural network must learn.

In the same spirit, we can modify the loss function by multiplying the residual $\mathcal{R}[u_\theta]$ with a spatially dependent function $\rho$, and construct the loss from the modified residual
\begin{equation}
\widetilde{\mathcal{L}}_{\text{pde}}(\theta)
=  \int_{\Omega}\left(\rho(x) \mathcal{R}[u_\theta](x)\right)^2\,dx \,.
\end{equation}
Such transformations can improve the numerical conditioning of the optimization problem or remove apparent singular factors arising from the PDE coordinate representation.
For example, coordinate-dependent rescalings have been used to regularize PINN residuals near coordinate singularities \cite{Okazaki2022,Tan2026}, while other works have rescaled or normalized PDE residuals to reduce large magnitude variations and improve training stability \cite{Dahlbudding2024,Tseneklidou2025}. Although \(\mathcal{R}[u]=0\) and \(\rho\mathcal{R}[u]=0\) are equivalent wherever $\rho\neq0$, the corresponding least-squares objectives are generally different, since the transformation changes the relative contribution of different regions of the domain to the training loss.

As mentioned in Section~\ref{sec:nn_formalism}, due to the nonlinear nature of neural networks, the training process of seeking the optimal parameters $\theta$ under the loss $\mathcal{L}(\theta)$ translates into a nonlinear optimization problem. This optimization is commonly performed using first-order gradient-based algorithms such as stochastic gradient descent (SGD), Adam, AdaGrad, and RMSProp. One of the most popular methods in deep learning applications is Adam~\cite{DBLP:journals/corr/KingmaB14}, which combines momentum-based updates with adaptive learning-rate scaling.
Quasi-Newton methods, such as Broyden--Fletcher--Goldfarb--Shanno (BFGS) and Broyden, present an alternative to first-order algorithms. These methods incorporate approximate curvature information through iterative estimations of the Hessian matrix, resulting in faster local convergence. For more details on these methods, see Ref.~\cite{nocedal_numerical_2006}.

Recent works~\cite{URBAN2025113656, kiyani2025optimizing,wang2025high} have introduced training strategies that combine these approaches: first-order methods during the early stages to explore the high-dimensional parameter space for $\theta$, and quasi-Newton methods such as L-BFGS or self-scaled Broyden updates to achieve higher precision in optimizing the loss.

\subsection{Error control}
\label{sec:pinns_convergence}

To ensure a correct approximation of the neural network to the PDE solution through PINNs, we must consider three sources of error: the optimization error, the generalization error, and the approximation error~\cite{Lu2021deepxde}. Before discussing these, we must clarify some notation and concepts.

For simplicity, we consider the weak imposition case, though this analysis also applies to the strong case. First, the loss function is rewritten to make explicit the dependence on the training points
\begin{equation}
\mathcal{L}(\theta;\mathcal{T}) = \mathcal{L}_{\text{pde}}(\theta;\mathcal{T}_{\text{pde}}) + \lambda\mathcal{L}_{\text{bc}}(\theta;\mathcal{T}_{\text{bc}})\,,
\end{equation}
where $\mathcal{T}_{\text{pde}}$ denotes the set of training points within the domain $\Omega$, $\mathcal{T}_{\text{bc}}$ denotes the set of training points over the boundary $\partial \Omega$, and $\mathcal{T}\coloneq\{\mathcal{T}_{\text{pde}}, \mathcal{T}_{\text{bc}}\}$. 

Second, we define $\mathcal{U}_{\theta}$ as the set of all neural network functions $u_{\theta}$ generated by varying $\theta$ given a fixed architecture (i.e., fixed number of layers, number of neurons, activation functions, etc).

We denote the best approximation $u_{\theta^*,\mathcal{T}}$ for a fixed set of training points as
\begin{equation}
u_{\theta^*,\mathcal{T}} \coloneq \argmin_{u_\theta \in \mathcal{U}_\theta}\mathcal{L}(\theta;\mathcal{T})\,,
\end{equation}
and the best approximation for the continuous loss
\begin{equation}
u_{\theta^*} \coloneq \argmin_{u_\theta \in \mathcal{U}_\theta}\left(\|\mathcal{R}[u_\theta]\|^{2}_{L^{2}(\Omega)} + \lambda \|\mathcal{R}_{\text{bc}}[u_\theta]\|^{2}_{L^{2}(\partial \Omega)}\right)\,.
\end{equation}
For simplicity, we assume the existence of these minimizers, although $\mathcal{U}_\theta$ might not be closed~\cite{PetRasVoiFOCM2021}. A more rigorous approach using quasi-minimizers can be found in~\cite{Shin2023error, Brevis2022neural}.

Then the approximation error between the current estimation $u_{\theta,\mathcal{T}}$ and the PDE solution $u$ can be bounded by (see, e.g., Ref.~\cite{Lu2021deepxde}):
\begin{equation}\label{eq:error_bounds}
\|u_{\theta,\mathcal{T}} - u\|\leq \underbrace{\|u_{\theta,\mathcal{T}} - u_{\theta^{*},\mathcal{T}}\|}_{\text{(optimization)}} + \underbrace{\|u_{\theta^{*},\mathcal{T}} - u_{\theta^{*}}\|}_{\text{(generalization)}} + \underbrace{\|u_{\theta^{*}} - u\|}_{\text{(approx.)}}.
\end{equation}
\begin{itemize}
    \item {\bf Optimization error}: Measures how close the optimization solver can drive the neural network to its best approximation $u_{\theta^*,\mathcal{T}}$. Different factors can influence this error, such as the selected algorithm, optimizer hyperparameter, and stopping criteria.
    \item {\bf Generalization error}: Measures how well the discretization of the integral can represent the continuous one (i.e., can be described as an integration error).
    Integration of neural networks, given their complex structure, has proven to be a challenging task~\cite{RIVERA2022114710}.
    This error is generated by the numerical integration approach (e.g., Monte Carlo) and by the strategy for selecting the set of collocation points (e.g., resampling $\mathcal{T}$ every few epochs or adaptive sampling strategies).
    \item {\bf Approximation error}: Measures how close the neural network set $\mathcal{U}_\theta$ is to the  exact solution $u$. This has been studied in several works, such as \cite{cybenko1989approx, HORNIK1991251,Pinkus_1999,NIPS2017_lu, YAROTSKY2017103}, and is known as the expressivity of the neural network.
\end{itemize}

Keeping these three errors under control is essential for proper training of the neural network and convergence to the correct solution.

\section{Implementation details}
\label{sec:code_implementation}
\pinndorama is implemented in Python using \texttt{JAX}, which provides automatic differentiation and just-in-time compilation for GPU execution. Automatic differentiation is used to compute both the spatial derivatives entering the differential-equation residuals and the gradients of the loss with respect to the network parameters.

Our implementation uses PINNs to solve problems of increasing complexity. We begin with the one-dimensional test problem in~\eqref{eq:1d-toy-pde}, which admits an exact solution and serves as a testbed for the techniques introduced here. We then solve the Hamiltonian constraint in~\eqref{eq:general-Hamiltonian} for axisymmetric and fully three-dimensional binary puncture configurations. The axisymmetric solver also includes a parametric implementation that takes the common spin of the two punctures as an additional network input. The puncture solvers draw their reference-metric and Bowen--York source expressions from \nrpy~\cite{Ruchlin:2017com}. This provides a direct connection to the coordinate systems and mathematical expressions used by \nrpyell~\cite{assumpcao_fast_2022}. In particular, we formulate the PINN residuals and sample the computational domain in the same native curvilinear coordinates used by these codes.

\subsection{Coordinate systems and sampling points} \label{sec:coordinates-systems}
A particular numerical challenge of the puncture initial data problem is that
the solution varies considerably near the punctures but decays slowly in the far
field. The distribution of collocation points can therefore have a substantial
effect on training, since the network must resolve both the near-puncture
structure and the asymptotic behavior over several orders of magnitude in
radius. To address this challenge, we use radially stretched coordinates on a truncated domain that provide comparatively dense sampling near the punctures while extending the computational domain far into the asymptotic region.
We follow the reference-metric infrastructure of \nrpy, in which curvilinear coordinate systems are expressed in terms of the logically Cartesian native coordinates $(\xxzero,\xxone,\xxtwo)$.

The one-dimensional problem uses \texttt{SinhSpherical} coordinates. Its
physical radius $r$ is related to the radial native coordinate
$\xxzero\in[0,1]$ by
\begin{equation}
    r(\xxzero)
    =r_{\max}\frac{\sinh(\xxzero/w)}{\sinh(1/w)}\, .
    \label{eq:sinh_spherical_map}
\end{equation}
Here $r_{\max}$ sets the physical radius of the outer boundary, while
$w>0$ controls the radial stretching. Smaller values of $w$ place a larger
fraction of the collocation points near the origin.

The two- and three-dimensional puncture problems use the \SinhSymTP
coordinates of Ref.~\cite{assumpcao_fast_2022}. We first define the
radial-like quantity
\begin{equation}
    \tilde{r}(\xxzero)
    =r_{\max}\frac{\sinh(\xxzero/w)}{\sinh(1/w)}\, .
\end{equation}
The transformation from the native coordinates to the Cartesian coordinates
$(x,y,z)$ is then
\begin{align}
    x & = \tilde{r}\sin(\xxone)\cos(\xxtwo)\, , \nonumber \\
    y & = \tilde{r}\sin(\xxone)\sin(\xxtwo)\, , \nonumber \\
    z & = \sqrt{\tilde{r}^{\,2}+b^2}\cos(\xxone)\, ,
    \label{eq:sinh_symtp_map}
\end{align}
where
$\xxzero\in[0,1]$, $\xxone\in[0,\pi]$, and
$\xxtwo\in[-\pi,\pi]$. The coordinate $\xxzero$ is radial-like,
$\xxone$ is polar-like, and $\xxtwo$ is azimuthal. The constant $b>0$
is the focal scale of the bipolar coordinate system. The ordinary spherical
radius associated with the Cartesian coordinates is
\begin{equation}
    r=\sqrt{x^2+y^2+z^2}
    =\sqrt{\tilde{r}^{\,2}+b^2\cos^2(\xxone)}\, .
\end{equation}
This mapping provides increased resolution near each puncture at $z=\pm b$ and
approaches a spherical coordinate system at large radius.
The axisymmetric solver uses the same \SinhSymTP mapping but omits the dependence on the
azimuthal coordinate $\xxtwo$.

For all calculations reported here, we set $r_{\max}=10^6$ and $w=0.07$.
These choices place the outer boundary well within the asymptotic region while
retaining dense sampling near the origin or punctures. The interior collocation points are constructed as
tensor products of cell-centered samples in each native coordinate direction. If
$\xx^{(a)}$ denotes the corresponding native coordinate, with lower and
upper bounds $\xx^{(a)}_{\min}$ and $\xx^{(a)}_{\max}$, its
cell-centered samples are
\begin{equation}
    \begin{split}
        \xx^{(a)}_j
        &=\xx^{(a)}_{\min}
        +\left(j+\frac{1}{2}\right)\Delta\xx^{(a)}\, ,\\
        \Delta\xx^{(a)}
        &=\frac{\xx^{(a)}_{\max}-\xx^{(a)}_{\min}}{N_a}\, ,
        \qquad j=0,\ldots,N_a-1.
    \end{split}
\end{equation}
Here, $N_a$ is the number of collocation points in direction $a$,
$\Delta\xx^{(a)}$ is the corresponding uniform coordinate spacing,
$j$ labels an individual point, and $\xx^{(a)}_j$ is the coordinate of
its $j$-th cell center. The complete interior collocation set is the
Cartesian product of these one-dimensional sample sets.

\subsection{Boundary conditions}
\label{sec:boundary_conditions}

The boundary terms entering the loss serve two distinct purposes. First, the outer boundary condition encodes the physical asymptotic behavior of the solution. Second, regularity and symmetry conditions ensure that the neural network represents a smooth, single-valued scalar field across coordinate boundaries. The form of these additional conditions depends on the coordinate system.

\subsubsection{Outer boundary condition}

At leading order, asymptotic flatness requires
\begin{align}
    u(r)   & = \frac{C}{r}+\mathcal{O}(r^{-2})\, , \nonumber \\
    \partial_r u(r) & = -\frac{C}{r^2}+\mathcal{O}(r^{-3})\, ,
\end{align}
where $C$ is not known in advance. Eliminating $C$ between these relations gives the first-order Robin residual
\begin{equation}
    \mathcal{B}_1[u]\equiv r \partial_r u+u=\mathcal{O}(r^{-2})\, .
    \label{eq:first_order_robin}
\end{equation}
At the finite outer computational boundary, we neglect the higher-order remainder and impose $\mathcal{B}_1[u]=0$.\footnote{In our implementation, the first-order residual is multiplied by $(1+r)$. This rescaling changes its relative weighting at large radius but does not change the boundary condition $\mathcal{B}_1[u]=0$.}

The boundary condition can be extended by one asymptotic order by retaining the next term,
\begin{align}
    u(r)   & = \frac{C}{r} + \frac{D}{r^2} + \mathcal{O}(r^{-3})\, , \nonumber \\
    \partial_r u(r) & = -\frac{C}{r^2} -\frac{2D}{r^3} +  \mathcal{O}(r^{-4})\, .
\end{align}
Applying the first-order operator to this expansion gives
\begin{equation}
    r \partial_r u+u=-\frac{D}{r^2}+\mathcal{O}(r^{-3})\, .
\end{equation}
Thus, although the first-order operator eliminates the unknown leading coefficient $C$, it does not eliminate the subleading coefficient $D$. For the 1D test problem, asymptotic matching of the field equation fixes $D=-2m$, as derived in Appendix~\ref{sec:appendix_second_order_robin}. Substituting this value yields the second-order Robin residual
\begin{equation}
\label{eq:second_order_robin}
    \mathcal{B}_2[u]\equiv
    r^3\left(\partial_r u+\frac{u}{r}\right)-2m
    =\mathcal{O}(r^{-1})\, .
\end{equation}
Because the coefficient $D=-2m$ is specific to the 1D test equation, we impose the second-order condition only for that problem.

We enforce the outer boundary condition either at the outer endpoint, $\xxzero=1$, or throughout an asymptotic radial region, including that endpoint. For the one-dimensional problem, we compare the first-order condition, $\mathcal{B}_1[u]=0$, imposed only at the outer endpoint, with the second-order condition, $\mathcal{B}_2[u]=0$, imposed at all sampled points satisfying $r>r_{\rm B}$. Here $r_{\rm B}=10^3$ in code units specifies the inner edge of the boundary-enforcement region. The two- and three-dimensional puncture problems use the first-order condition imposed only on the outer surface $\xxzero=1$.

\subsubsection{Coordinate identifications} \label{sec:coord_identifications}
The coordinate identification losses described below are used only by the 2D
and 3D puncture solvers.
Because the collocation points are expressed in a curvilinear coordinate
system, distinct native coordinate labels
$(\xxzero,\xxone,\xxtwo)$ may represent the same Cartesian location. Therefore, we
may optionally supplement the loss with terms requiring the scalar field to agree
at such identified points.

Our construction adapts the inner boundary condition introduced in
Ref.~\cite{Ruchlin:2017com}. Let
$\mathbf{X}(\xxzero,\xxone,\xxtwo)=(x,y,z)$ denote the \SinhSymTP map defined
by~\eqref{eq:sinh_symtp_map}. Because $u$ is a scalar, two native coordinate
points $\mathbf{q}$ and $\mathbf{q}'$ must satisfy
$u(\mathbf{q})=u(\mathbf{q}')$ whenever
$\mathbf{X}(\mathbf{q})=\mathbf{X}(\mathbf{q}')$. Applying the \SinhSymTP map
gives the five inner boundary point identifications
\begin{subequations}
    \label{eq:sinh_symtp_inner_mappings}
    \begin{align}
        \mathbf{X}(-\xxzero,\xxone,\xxtwo) &= \mathbf{X}(\xxzero,\xxone,\pi+\xxtwo)\, , \label{eq:sinh_symtp_radial_mapping}\\
        \mathbf{X}(\xxzero,-\xxone,\xxtwo) &= \mathbf{X}(\xxzero,\xxone,\pi+\xxtwo)\, , \label{eq:sinh_symtp_north_polar_mapping}\\
        \mathbf{X}(\xxzero,\pi+\xxone,\xxtwo) &= \mathbf{X}(\xxzero,\pi-\xxone,\pi+\xxtwo)\, , \label{eq:sinh_symtp_south_polar_mapping}\\
        \mathbf{X}(\xxzero,\xxone,-\pi-\xxtwo) &= \mathbf{X}(\xxzero,\xxone,\pi-\xxtwo)\, , \label{eq:sinh_symtp_lower_azimuthal_mapping}\\
        \mathbf{X}(\xxzero,\xxone,\pi+\xxtwo) &= \mathbf{X}(\xxzero,\xxone,-\pi+\xxtwo)\, . \label{eq:sinh_symtp_upper_azimuthal_mapping}
    \end{align}
\end{subequations}

The first rule, \eqref{eq:sinh_symtp_radial_mapping}, identifies equivalent
points across the $\xxzero=0$ boundary, which maps to the focal line segment
$(x,y,z)=(0,0,b\cos\xxone)$. The polar identifications are given by
\Crefrange{eq:sinh_symtp_north_polar_mapping}
{eq:sinh_symtp_south_polar_mapping}, while
\Crefrange{eq:sinh_symtp_lower_azimuthal_mapping}
{eq:sinh_symtp_upper_azimuthal_mapping} express azimuthal periodicity. These
relations place no symmetry restriction on the physical solution $u$; they
require only that coordinate labels representing the same Cartesian point
yield the same scalar value.

In practice, our implementation does not enforce all five identifications in~\eqsref{eq:sinh_symtp_inner_mappings}. The 1D solver does not require terms for coordinate identification, since regularity at the origin is imposed directly through the condition $\partial_r u(0)=0$. The 2D solver enforces the two polar identifications, while the 3D solver additionally enforces azimuthal periodicity. Neither puncture solver includes a loss term associated with the radial-like identification across $\xxzero=0$.

For the axisymmetric 2D solver, the code enforces the matching conditions
\begin{align}
    u(\xxzero,-\delta) &= u(\xxzero,\delta)\, , \nonumber\\
    u(\xxzero,\pi+\delta) &= u(\xxzero,\pi-\delta)\, ,
    \label{eq:2d_polar_identification}
\end{align}
which correspond to the axisymmetric versions of~\Crefrange{eq:sinh_symtp_north_polar_mapping}{eq:sinh_symtp_south_polar_mapping}. These identifications are enforced by adding the following contribution to the loss:
\begin{align}
    \mathcal{L}_{\mathrm{polar}}^{\mathrm{2D}}
    &=\frac{1}{2}\left\langle\left[u(\xxzero,-\delta)-u(\xxzero,\delta)\right]^2\right\rangle \nonumber \\
    &+\frac{1}{2}\left\langle\left[u(\xxzero,\pi+\delta)-u(\xxzero,\pi-\delta)\right]^2\right\rangle \,.
    \label{eq:2d_polar_loss}
\end{align}
Here $\langle\cdot\rangle$ denotes the arithmetic mean over the corresponding
sample set. The $\xxzero$ coordinate is sampled at cell-centered points
spanning the interval $[0,1]$, while $\delta>0$ denotes the native coordinate
distance from either polar boundary.\footnote{The polar matching region uses
cell-centered offsets satisfying $0<\delta<0.1$.}

For the 3D solver, the network depends explicitly on the azimuthal coordinate,
so crossing a polar axis requires an accompanying shift of $\xxtwo$ by $\pi$.
Because $\xxtwo+\pi$ may fall outside the native interval $[-\pi,\pi)$, we
use azimuthal periodicity to map the shifted angle back into this interval:
\begin{equation}
    \xxtwo_\star\equiv
    \begin{cases}
        \xxtwo+\pi, & -\pi\leq\xxtwo<0\, ,\\
        \xxtwo-\pi, & 0\leq\xxtwo<\pi\, .
    \end{cases}
    \label{eq:3d_axis_shifted_xx2}
\end{equation}
Thus, $\xxtwo_\star$ and $\xxtwo+\pi$ represent the same azimuthal angle.
For each sampled $(\xxzero,\delta,\xxtwo)$, the code imposes the rules in~\Crefrange{eq:sinh_symtp_north_polar_mapping}{eq:sinh_symtp_south_polar_mapping} as
\begin{align}
    u(\xxzero,-\delta,\xxtwo_\star)
    &=u(\xxzero,\delta,\xxtwo)\, , \nonumber\\
    u(\xxzero,\pi+\delta,\xxtwo_\star)
    &=u(\xxzero,\pi-\delta,\xxtwo)\, ,
    \label{eq:3d_polar_identification}
\end{align}
by including the following contribution to the loss
\begin{align}
    \mathcal{L}_{\mathrm{polar}}^{\mathrm{3D}}
    &=\frac{1}{2}\left\langle
    \left[u(\xxzero,-\delta,\xxtwo_\star)
    -u(\xxzero,\delta,\xxtwo)\right]^2
    \right\rangle \nonumber\\
    &+\frac{1}{2}\left\langle
    \left[u(\xxzero,\pi+\delta,\xxtwo_\star)
    -u(\xxzero,\pi-\delta,\xxtwo)\right]^2
    \right\rangle \,,
    \label{eq:3d_polar_loss}
\end{align}
with both $\xxzero$ and $\xxtwo$ covering the whole domain, and $\delta$ sampled in the same small region as in the 2D case.

The 3D solver also enforces periodicity across the azimuthal boundary
by condensing the effects of~\Crefrange{eq:sinh_symtp_lower_azimuthal_mapping}{eq:sinh_symtp_upper_azimuthal_mapping} into a single loss contribution:
\begin{equation}
    \mathcal{L}_{\rm azim}^{\mathrm{3D}}
    =\left\langle
    \left[u(\xxzero,\xxone,\pi)
    -u(\xxzero,\xxone,-\pi)\right]^2
    \right\rangle .
    \label{eq:3d_xx2_periodicity}
\end{equation}

\subsection{Loss function and output rescaling}

Before constructing the loss function, we rescale the raw neural network output using a fixed radial prefactor,
\begin{equation}
    u_\theta=\frac{nn_\theta}{\sqrt{1+r^2}}\,.
\end{equation}
This factor incorporates the expected $1/r$ asymptotic scaling while remaining finite at the origin. It does not embed an approximate analytical solution: the network must still learn the solution from the governing equation and boundary conditions. The radial coordinate $r$ is defined by~\eqref{eq:sinh_spherical_map} for the one-dimensional problem and computed from~\eqsref{eq:sinh_symtp_map} for the puncture problems.

For all solvers, the interior loss is the arithmetic mean of the squared regularized residual over $N$ collocation points,
\begin{equation}
    \mathcal{L}_{\mathrm{pde}}(\theta)
    =\frac{1}{N}\sum_{i=1}^{N}
    \left[\hat{\mathcal{R}}[u_\theta]\right]_i^2\,.
\end{equation}
Here and below, the subscript $i$ denotes evaluation at the corresponding sample point. Spatial derivatives of $u_\theta$ are computed using automatic differentiation. The residual and additional loss terms are specified separately for the one-dimensional and puncture problems.

For the one-dimensional test problem, the regularized residual is
\begin{equation}\label{eqn:1d_toy_regulatization}
    \hat{\mathcal{R}}_{\mathrm{1D}}[u]
    =r^2\mathcal{R}_{\mathrm{1D}}[u]\,,
\end{equation}
where $\mathcal{R}_{\mathrm{1D}}$ is the left-hand side of~\eqref{eq:1d-toy-pde}, including the manufactured source $A(r)$. Regularity at the origin is imposed through
\begin{equation}
    \mathcal{L}_{\mathrm{origin}}(\theta)
    =\left|\left.\partial_r u_\theta\right|_{r=0}\right|^2\,.
\end{equation}
The first-order Robin experiment uses the boundary loss
\begin{equation}
    \mathcal{L}_{\mathrm{bc}}(\theta)
    =\frac{1}{N_{\mathrm{bc}}}\sum_{i=1}^{N_{\mathrm{bc}}}
    \left[(1+r)\mathcal{B}_1[u_\theta]\right]_i^2\,,
\end{equation}
evaluated only at the outer endpoint. The positive factor $(1+r)$ leaves the boundary condition unchanged while counteracting the suppression of its loss contribution at large radii. The second-order experiment instead uses
\begin{equation}
    \mathcal{L}_{\mathrm{bc}}(\theta)
    =\frac{1}{N_{\mathrm{bc}}}\sum_{i=1}^{N_{\mathrm{bc}}}
    \left[\mathcal{B}_2[u_\theta]\right]_i^2\,,
\end{equation}
evaluated at all sampled interior points satisfying $r>r_{\mathrm B}$ together with the outer endpoint, without an additional $(1+r)$ factor. The operators $\mathcal{B}_1$ and $\mathcal{B}_2$ are defined in~\eqref{eq:first_order_robin} and~\eqref{eq:second_order_robin}, respectively. For either boundary condition, the total one-dimensional loss is
\begin{equation}
    \mathcal{L}_{\mathrm{tot}}^{1\mathrm{D}}(\theta)
    =\mathcal{L}_{\mathrm{pde}}(\theta)
    +\mathcal{L}_{\mathrm{bc}}(\theta)
    +\mathcal{L}_{\mathrm{origin}}(\theta)\,.
\end{equation}

For the two- and three-dimensional puncture problems, let $\hat{\gamma}_{ij}$ denote the flat three-dimensional metric expressed in the native coordinates, and let $\hat{\gamma}=\det\hat{\gamma}_{ij}$ denote its coordinate-dependent determinant. The Hamiltonian-constraint residual is
\begin{equation}\label{eq: general residual}
    \mathcal{R}[u]
    =\frac{1}{\sqrt{\hat{\gamma}}}\partial_i\left(\sqrt{\hat{\gamma}}\hat{\gamma}^{ij}\partial_j u\right)
    +\frac18\frac{\tilde{A}_{ij}\tilde{A}^{ij}}{(\psi_{\rm singular}+u)^7}\,.
\end{equation}
The axisymmetric problem uses the same metric and residual, with vanishing azimuthal derivatives of $u$. Both puncture solvers use the regularized residual
\begin{align}
    \hat{\mathcal{R}}[u]
    &=\sqrt{\hat{\gamma}}\mathcal{R}[u]\nonumber\\
    &=\partial_i\left(\sqrt{\hat{\gamma}}\hat{\gamma}^{ij}\partial_j u\right)
    +\frac{\sqrt{\hat{\gamma}}}{8}\frac{\tilde{A}_{ij}\tilde{A}^{ij}}{(\psi_{\rm singular}+u)^7}\,.
\end{align}
Their boundary loss uses the scaled first-order expression above, evaluated on the outer surface $\xxzero=1$. The coordinate-identification terms defined in Sec.~\ref{sec:coord_identifications} complete the losses:
\begin{align}
    \mathcal{L}_{\mathrm{tot}}^{2\mathrm{D}}(\theta)
    &=\mathcal{L}_{\mathrm{pde}}(\theta)+\mathcal{L}_{\mathrm{bc}}(\theta)
    +\mathcal{L}_{\mathrm{polar}}^{2\mathrm{D}}(\theta)\,,\\
    \mathcal{L}_{\mathrm{tot}}^{3\mathrm{D}}(\theta)
    &=\mathcal{L}_{\mathrm{pde}}(\theta)+\mathcal{L}_{\mathrm{bc}}(\theta)\nonumber\\
    &\quad+\mathcal{L}_{\mathrm{polar}}^{3\mathrm{D}}(\theta)
    +\mathcal{L}_{\mathrm{azim}}^{3\mathrm{D}}(\theta)\,.
\end{align}
All component multipliers are unity in the reported configurations, corresponding to $\lambda=1$ in the earlier notation.

\subsection{Training strategy} \label{sec:training-strategy}

Optimization is performed in two stages. The neural network is initially trained using Adam, whose adaptive first-order updates provide an effective initial reduction of the loss from randomly initialized parameters. The resulting parameters are then passed to the self-scaled Broyden (SSBroyden) quasi-Newton optimizer, which uses approximate curvature information to refine the solution \cite{kiyani2025optimizing, bioli2026ssBroyden}. As demonstrated below, this second stage can reduce both the training loss and the relative $L^2$ error by several orders of magnitude. We use the SSBroyden implementation provided as an add-on to the \texttt{Optimistix} library in \texttt{JAX}, including GPU support \cite{bioli2026ssBroyden}.

The initial Adam stage uses a fixed number of $1500$ steps for the one-dimensional, axisymmetric, and parametric problems, and $10000$ steps for the initial three-dimensional training. The SSBroyden training stage terminates when both of the following conditions are satisfied:
\begin{align}
\max_i
\frac{\left|\theta_i^{(k+1)}-\theta_i^{(k)}\right|}
{\epsilon_{\mathrm{abs}}+
 \epsilon_{\mathrm{rel}}\left|\theta_i^{(k+1)}\right|}
&<1,
\\
\frac{\left|\mathcal{L}^{(k+1)}-\mathcal{L}^{(k)}\right|}
{\epsilon_{\mathrm{abs}}+
 \epsilon_{\mathrm{rel}}\left|\mathcal{L}^{(k+1)}\right|}
&<1,
\end{align}
where $\theta_i^{(k)}$ is the $i$-th trainable parameter and $\mathcal{L}^{(k)}$ is the total training loss after the optimization step $k$. In all our runs, we set $\epsilon_{\mathrm{abs}}=\epsilon_{\mathrm{rel}}=10^{-12}$. Otherwise, SSBroyden terminates after a maximum of $10000$ steps for the one-dimensional problem or $50000$ steps for the puncture problems.

\section{Results}
\label{sec:results}

We validate the method against an analytic one-dimensional solution, then consider axisymmetric and fully three-dimensional binary puncture initial data. Finally, we train a parametric network and demonstrate its use as an initial guess for hyperbolic relaxation. All networks have $4$ hidden layers with $40$ neurons each and $\tanh$ activations, using Xavier initialization for the one-dimensional problem and He initialization for the puncture problems~\cite{glorot2010understanding,He2015}.

Neural networks were trained on NVIDIA A100 GPUs on the Tursa system using double-precision arithmetic. Each training run used a single GPU, with four independent runs executed concurrently on each compute node.

To measure the accuracy of the trained PINNs, we define the relative $L^2$ error as
\begin{equation}
\epsilon_{L^2}
=\left[
\frac{\sum_i w_i |u_{\theta,i}-u_{{\rm ref},i}|^2}
     {\sum_i w_i |u_{{\rm ref},i}|^2}
\right]^{1/2} \,,
\label{eq:relative_l2_error}
\end{equation}
where $u_{\theta,i}$ and $u_{{\rm ref},i}$ denote the neural-network and reference solutions evaluated at the same point. We use $w_i=1$ for the one-dimensional problem and $w_i=\sqrt{\hat{\gamma}(x_i)}$ for the puncture problems. In the latter, the flat-metric volume factor is evaluated at the three-dimensional spatial point $x_i=(\xxzero,\xxone,\xxtwo)_i$, expressed in native curvilinear coordinates.

The one-dimensional reference is the exact analytic solution. For both the axisymmetric and fully three-dimensional puncture problems, the \nrpyell reference solutions are generated on $256\times256\times24$ grids and evaluated using tensor-product Lagrange interpolation with nine points per coordinate direction. At comparable resolutions, calibration tests against \TwoPunctures demonstrated relative agreement of order $10^{-10}$~\cite{assumpcao_fast_2022}, below the PINN errors reported here, which reach order $10^{-8}$ in axisymmetry and a few times $10^{-5}$ in three dimensions.

\subsection{One-dimensional validation}
\label{sec:one_dimensional_results}

We first consider the nonlinear problem in \eqref{eq:1d-toy-pde}, using its analytic solution as the reference. We compare first-order Robin boundary conditions imposed at $r_{\max}=10^6$ with second-order boundary conditions imposed over $r>10^3$, including the outer endpoint, as described in Sec.~\ref{sec:boundary_conditions}. In both cases, we also impose regularity at the origin, $\partial_r u(0)=0$. The $L^2$ norms of relative errors are evaluated at 400 uniformly spaced cell centers in $\xxzero$, which map to nonuniformly spaced physical radii. The training of the neural networks and radial profile plots were performed with resolution $\Nzero=96$.
All displayed runs use the same seed for random initialization of the neural net parameters. The dependence on seed choice is discussed in Appendix~\ref{sec:appendix_seed_dependence}.

Figure~\ref{fig:1d_radial_relative_error} shows the pointwise relative errors for the two outer boundary conditions. Near $r=1$, the first-order condition yields errors of order $10^{-7}$, compared with $10^{-6}$ for the second-order condition, and remains more accurate up to approximately $r=100$. At the outermost evaluation point, however, its relative error reaches order $10^{-1}$, while the error of the second-order solution is around $10^{-5}$. Because the analytic solution decays as $r^{-1}$, far-field residuals contribute relatively little to the total loss. Imposing the second-order boundary condition over $r>10^3$ therefore places greater emphasis on accuracy in the outer region. Consequently, the second-order condition improves exterior accuracy at the cost of a modest increase in interior error.

\begin{figure}[h]
\centering
\includegraphics[width=\linewidth]{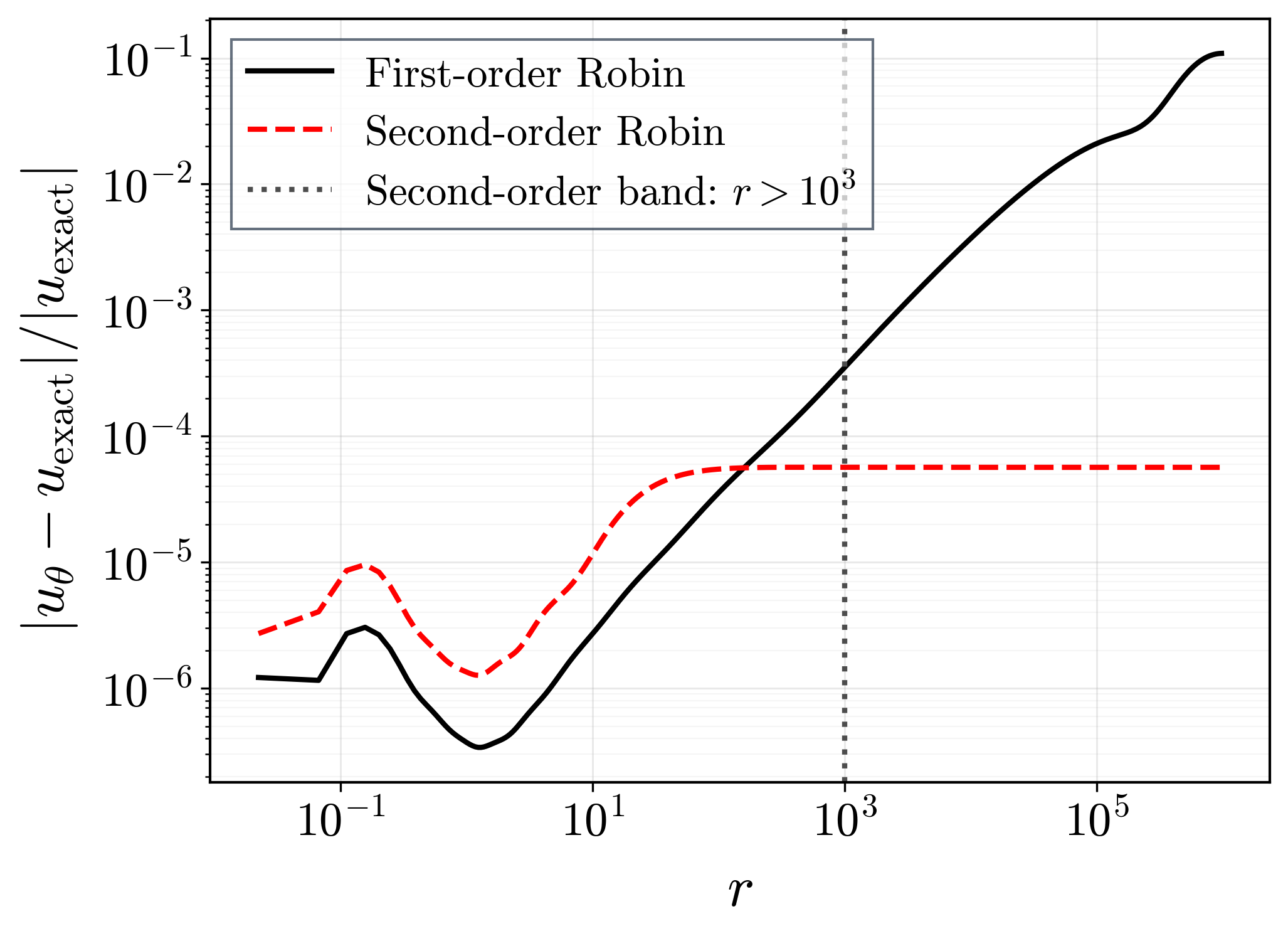}
\caption{Pointwise relative errors for the one-dimensional problem, with both networks trained at $\Nzero=96$. The vertical line marks the inner edge of the second-order boundary region, $r=10^3$.}
\label{fig:1d_radial_relative_error}
\end{figure}

\begin{figure}[h]
\centering
\includegraphics[width=\linewidth]{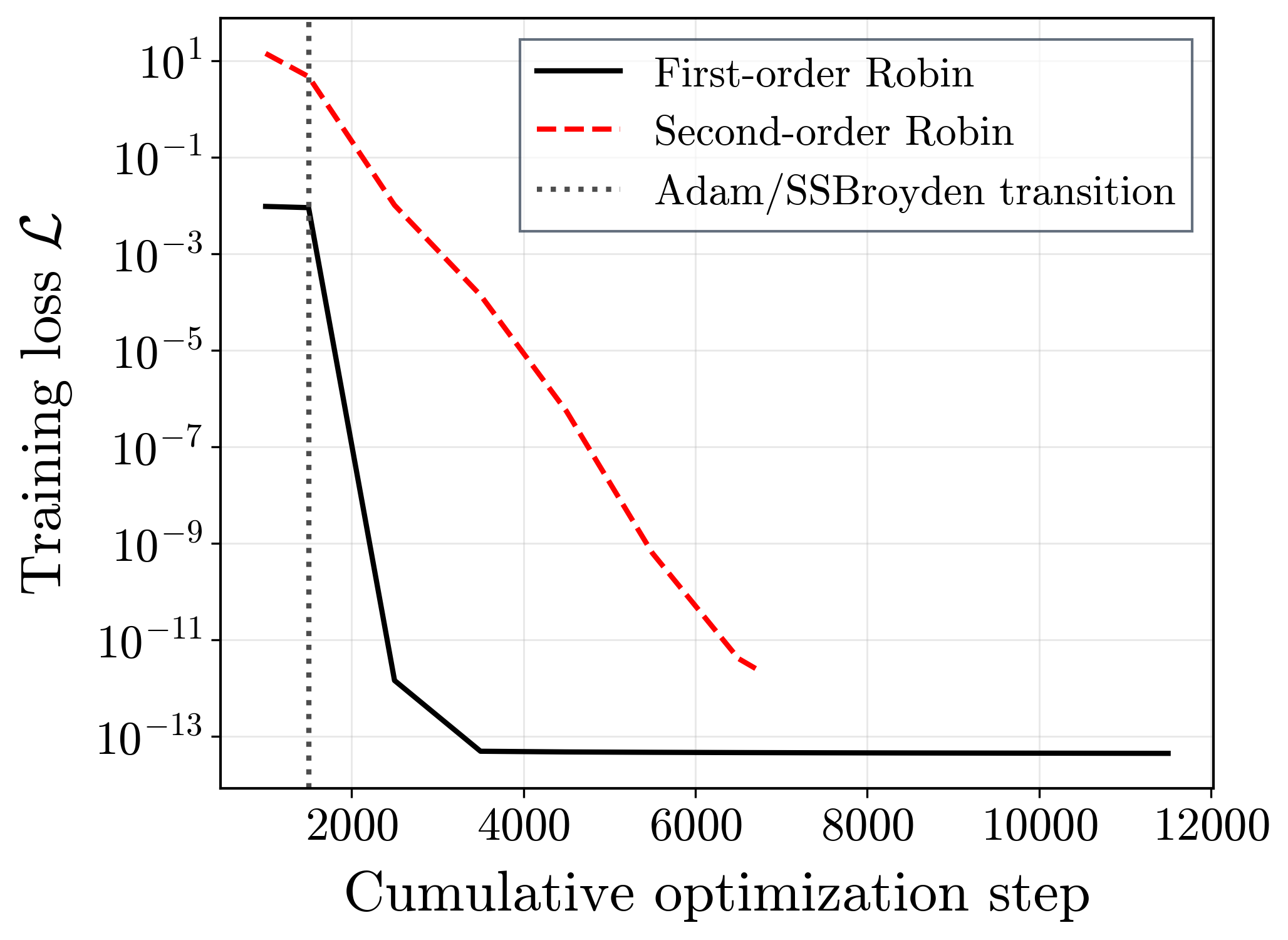}\par\smallskip
\includegraphics[width=\linewidth]{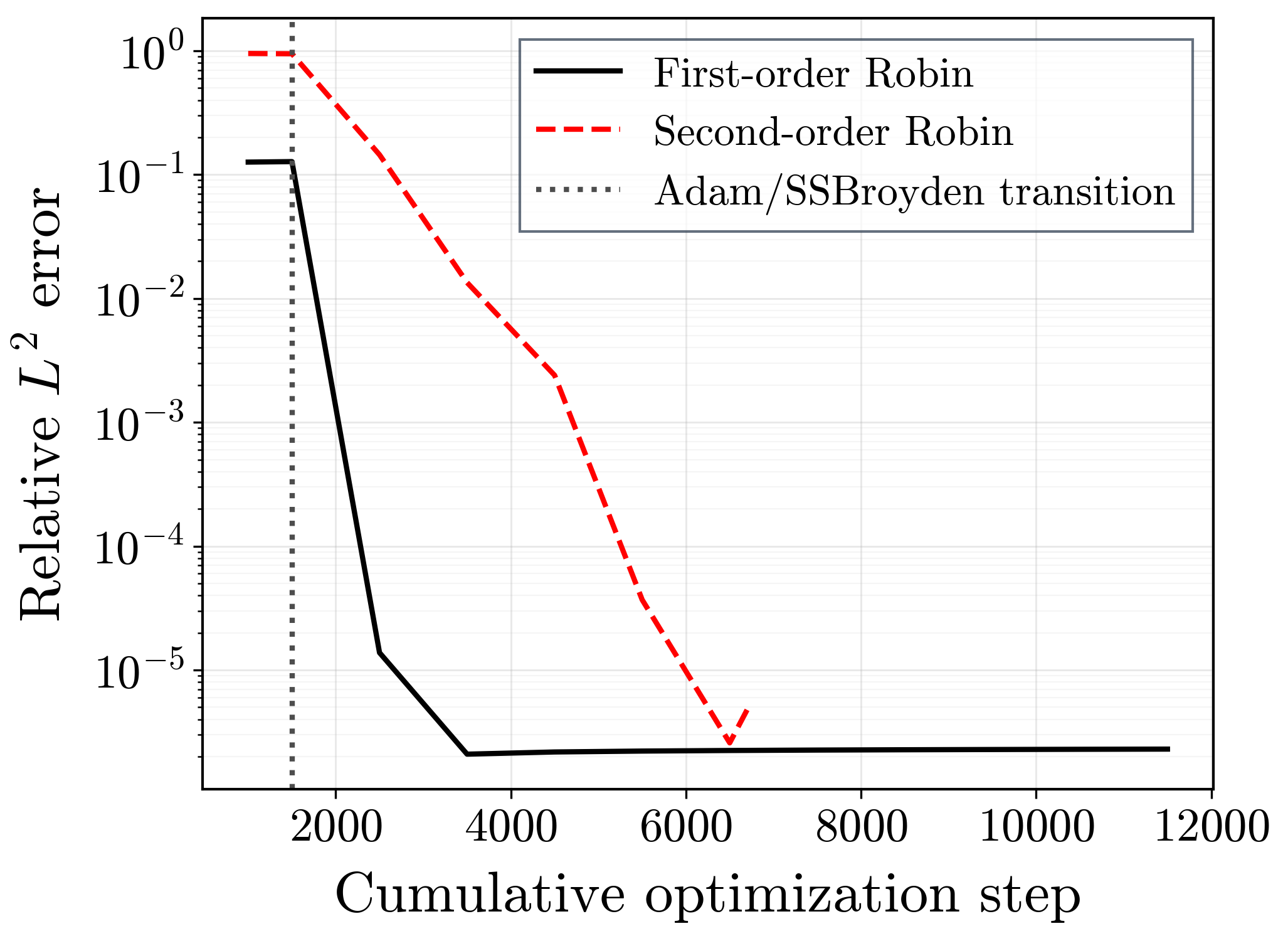}
\caption{Training loss (top) and relative $L^2$ error (bottom) for resolution $\Nzero=96$. Solid black and dashed red curves show first- and second-order Robin conditions, respectively. Dotted vertical lines mark the transition from Adam to SSBroyden training at step 1500.}
\label{fig:1d_training_history}
\end{figure}

Figure~\ref{fig:1d_training_history} shows the training loss and relative $L^2$ error as functions of cumulative optimization step. The Adam stage provides a small initial reduction in loss, and the relative errors remain approximately $0.1$ and $1$ for the first- and second-order conditions, respectively. SSBroyden subsequently reduces both to order $10^{-6}$. Further refinement with Adam would require additional optimization steps; in our tests, however, longer initial Adam training did not necessarily shorten the subsequent SSBroyden stage.

Each of the runs approaches its final accuracy at different rates. The first-order training reaches a relative error of order $10^{-6}$ by cumulative step 3,500, whereas the second-order run reaches this level by step 6,500. Nevertheless, the second-order run satisfies the stopping criterion after about 5,200 SSBroyden steps, or 6,700 cumulative steps, while the first-order run reaches the limit of 10,000 SSBroyden steps. Termination requires sufficiently small changes in both the network parameters and the loss, as specified in Sec.~\ref{sec:training-strategy}. Due to the conservative choice of stopping parameters, namely $\epsilon_{\mathrm{abs}}=\epsilon_{\mathrm{rel}}=10^{-12}$, the plateau in the first-order version of the loss did not lead to a stop in training.

Another important observation is that both the loss and the relative errors generally decrease as training progresses. In the context of the discussion in Sec.~\ref{sec:pinns_convergence}, the two panels in Figure~\ref{fig:1d_training_history} show consistent trends in the optimization and approximation errors. The only exception is a small increase in the relative $L^2$ error for the second-order boundary condition a few steps before the stopping criterion is satisfied. Identifying an optimum stopping criterion remains a subject for future investigation.

\begin{figure}[h]
\centering
\includegraphics[width=\linewidth]{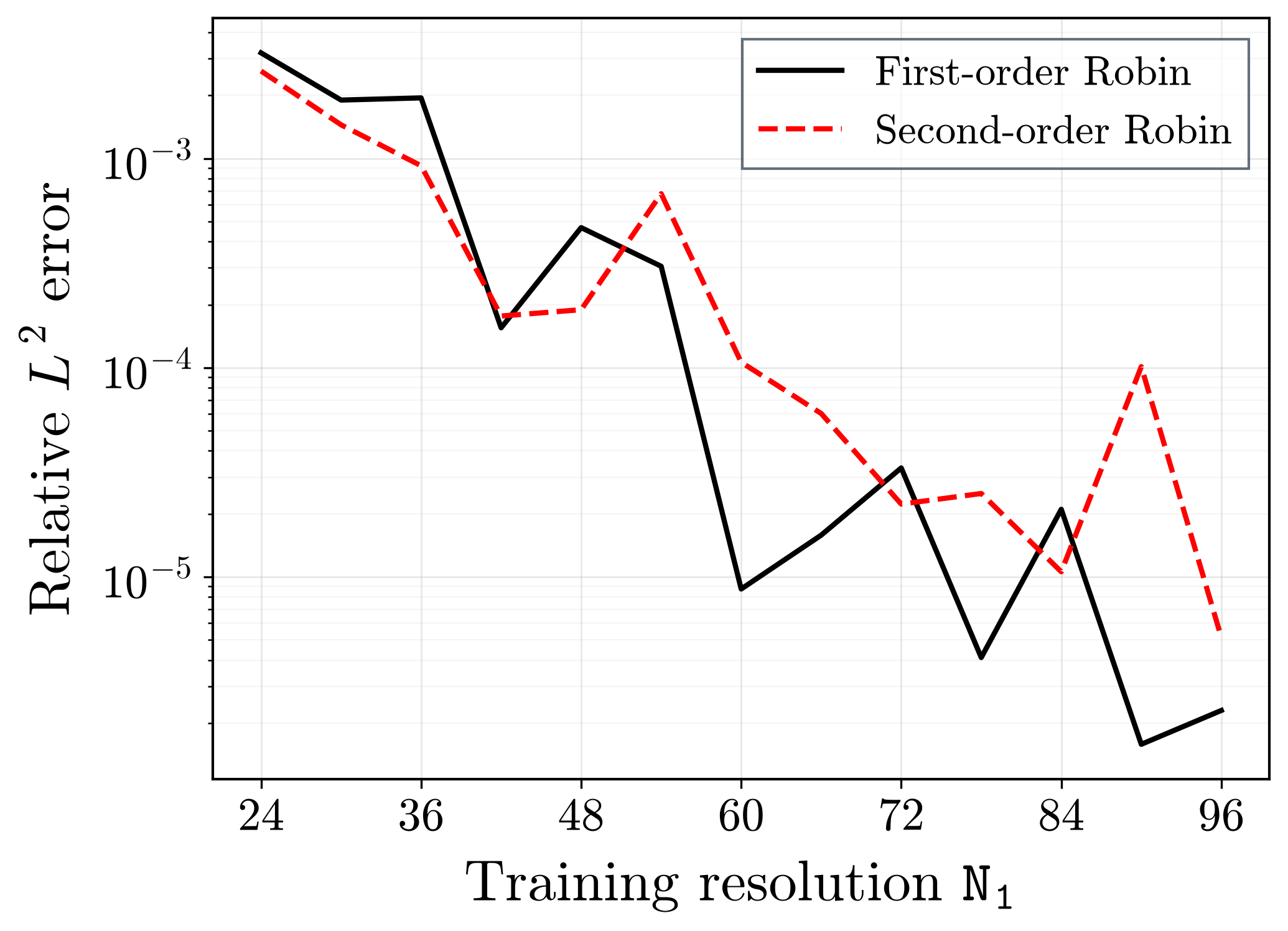}
\caption{Relative $L^2$ error versus training resolution, $\Nzero=24,30,\ldots,96$, for the final best-loss networks. Both conditions use seed 518 and the same independent evaluation grid.}
\label{fig:1d_convergence}
\end{figure}

Increasing the collocation-point resolution from $\Nzero=24$ to $\Nzero=96$ reduces the relative error from order $10^{-3}$ to order $10^{-6}$ for both conditions (Figure~\ref{fig:1d_convergence}). This improvement is substantial but non-monotonic. As discussed in Sec.~\ref{sec:pinns_convergence}, changing the collocation set modifies the discrete objective, while the fixed architecture and finite number of optimization steps continue to limit the attainable accuracy. Further discussion on the convergence of this PINN is presented in Appendix~\ref{sec:appendix_seed_dependence}.

\subsection{Axisymmetric binary-puncture initial data}
\label{sec:axisymmetric_2d_results}

Having validated our training procedure on a test problem, we now solve~\eqref{eq:general-Hamiltonian} for the non-singular correction $u$ to the conformal factor. We consider two punctures at $z_\pm=\pm2.5M$, with equal bare masses $m_+=m_-=M/2$\footnote{Here, $M=m_++m_-$, and we set $M=1$ in code units.} and vanishing linear momenta, $\mathbf{P}_\pm=0$. The subscripts $+$ and $-$ identify the positive- and negative-$z$ punctures, respectively. Their spins are $\mathbf{S}_{+}=(0,0,0.2M^2)$ and $\mathbf{S}_{-}=(0,0,-0.15M^2)$. This configuration is axially symmetric, but the unequal spin magnitudes break reflection symmetry across the equatorial plane. The correction can therefore be expressed as $u=u(\xxzero,\xxone)$.

We set the focal parameter to $b=2.5M$, matching the puncture positions, and impose the first-order Robin condition at the outer boundary. All runs presented here use equal resolutions in both coordinates, $\Nzero=\None$, and the same random seed for initializing the neural network parameters. The relative $L^2$ errors are evaluated on an independent $760\times760$ grid in native coordinates, restricted to $r<7.5M$.

\begin{figure}[h]
\centering
\includegraphics[width=\linewidth]{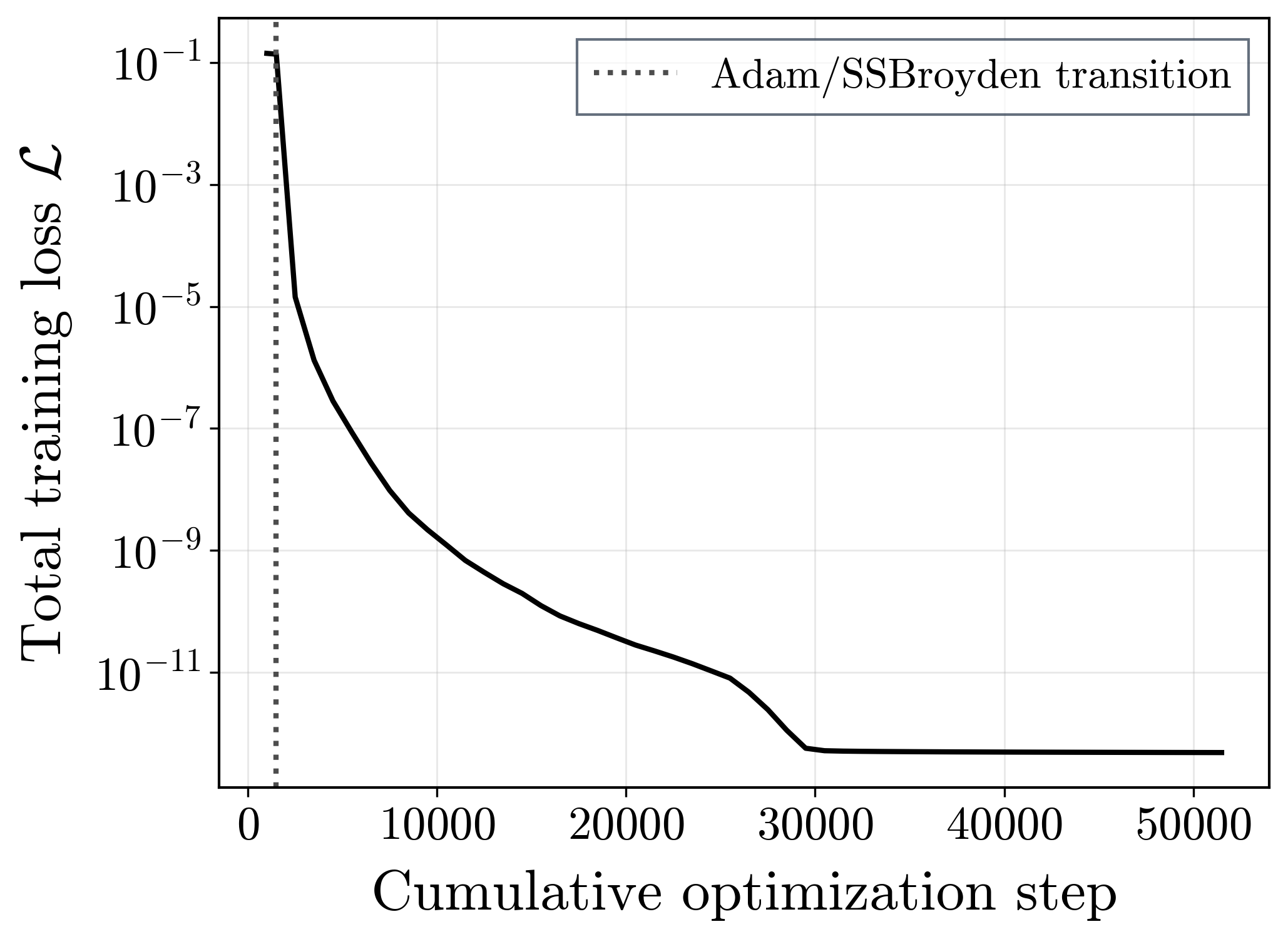}\par\smallskip
\includegraphics[width=\linewidth]{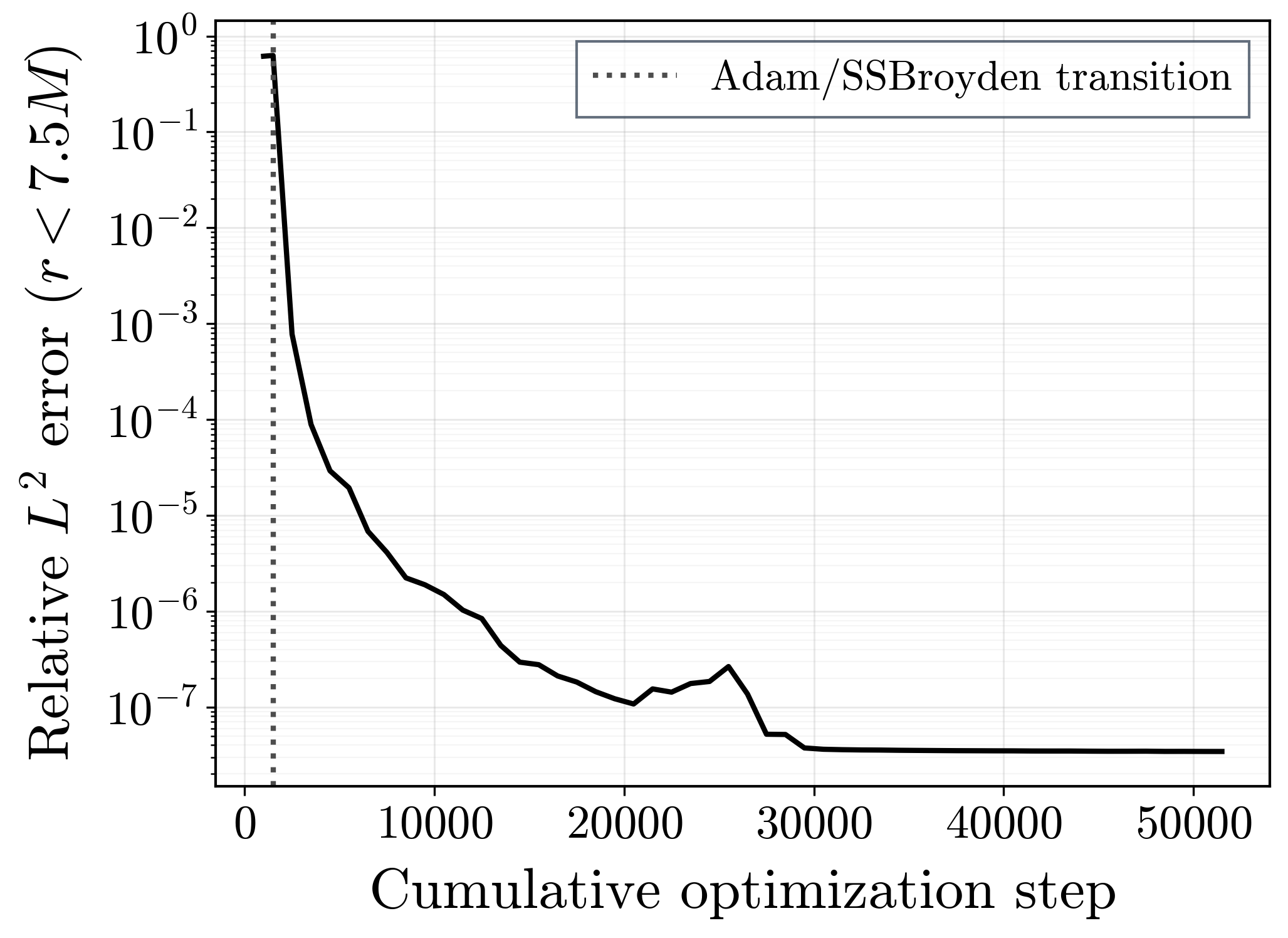}
\caption{Total training loss (top) and relative $L^2$ error inside $r<7.5M$ (bottom) for the $380\times380$ calculation, evaluated at saved network snapshots. The dotted vertical lines mark the Adam--SSBroyden transition at step 1500.}
\label{fig:2d_training_history}
\end{figure}

Figure~\ref{fig:2d_training_history} shows the training loss and relative $L^2$ error as functions of the cumulative number of optimization steps for the training resolution $(\Nzero,\None)=(380,380)$. The relative error remains of order unity at the end of the Adam stage, while subsequent optimization with SSBroyden reduces it to order $10^{-8}$ and the total training loss to below $10^{-12}$. Both diagnostics generally decrease throughout this refinement, although the relative error briefly increases near cumulative step 25,000 while the loss continues to decrease. After approximately 30,000 steps, both the relative error and loss plateau, yet the stopping criterion remains unsatisfied. The decrease in training loss is therefore generally accompanied by a decrease in the relative $L^2$ error. However, the stopping criterion remains overly conservative, as also observed in the one-dimensional problem.

\begin{figure}[h]
\centering
\includegraphics[width=\linewidth]{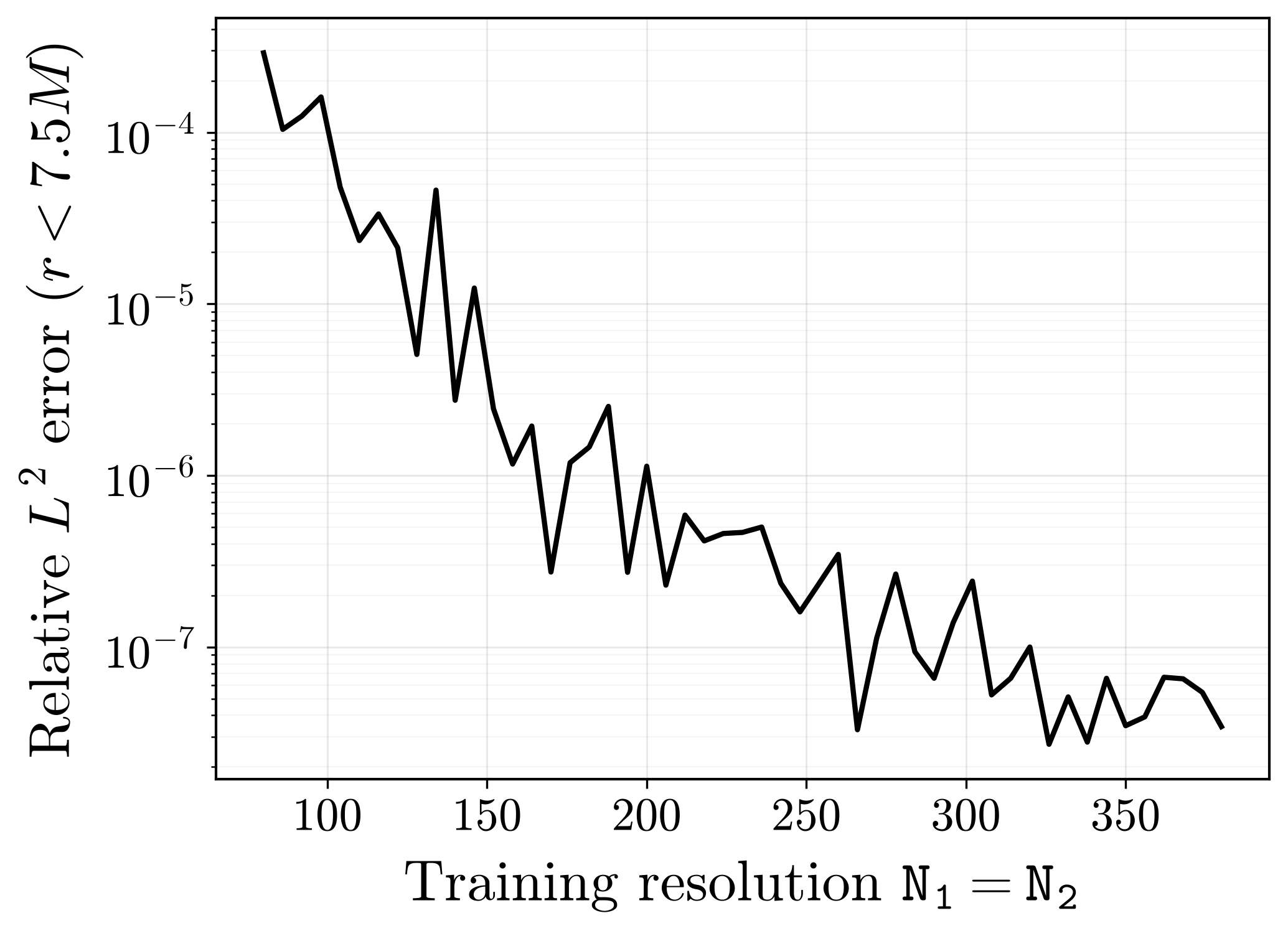}\par\smallskip
\includegraphics[width=\linewidth]{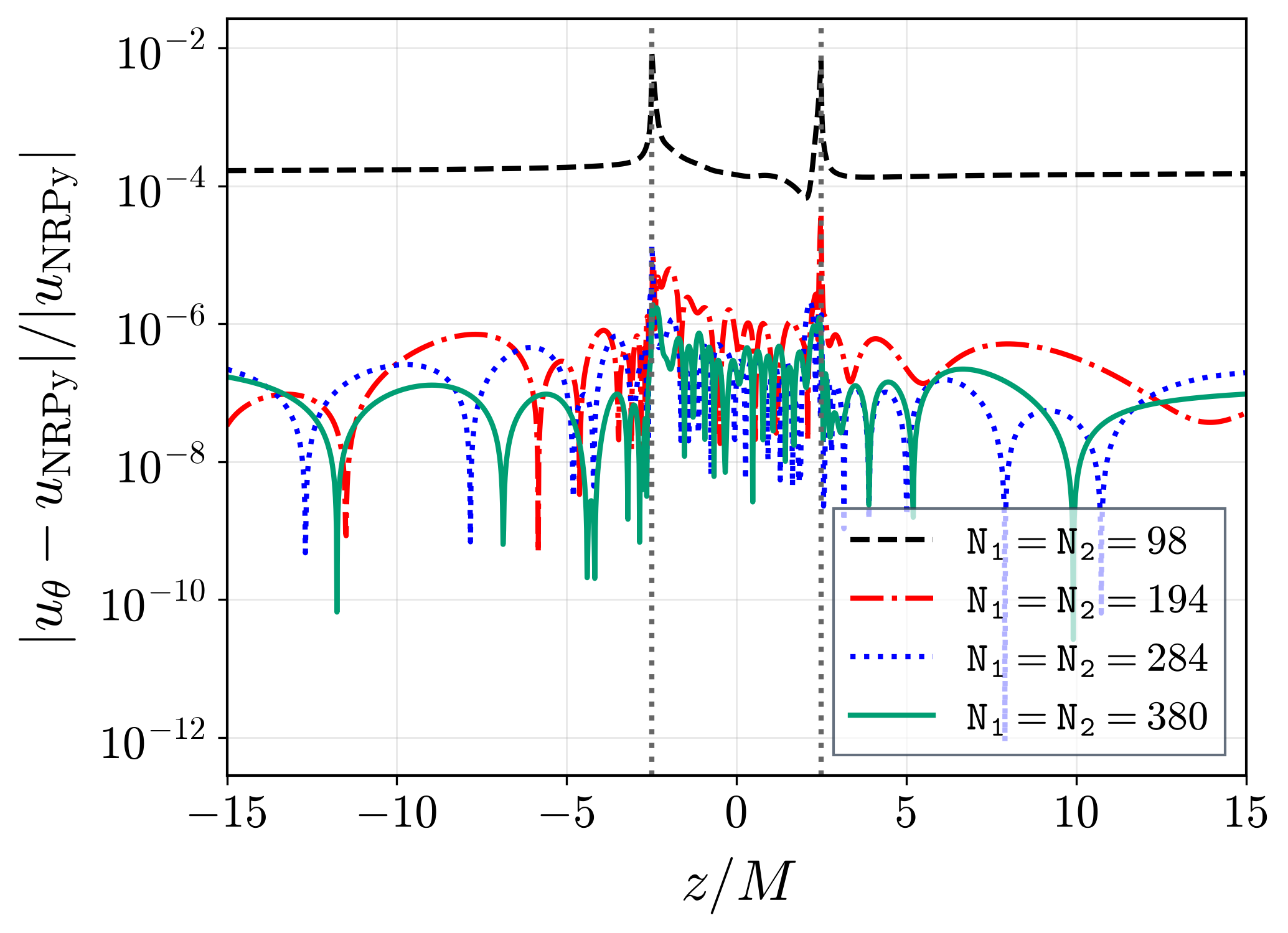}
\caption{Resolution dependence of the axisymmetric solution error. Top: relative $L^2$ error versus training resolution, $\Nzero=\None=80,86,\ldots,380$. Bottom: pointwise relative errors for $\Nzero=\None=98$, 194, 284, and 380 along the $z$ axis. The dotted vertical lines mark the puncture locations.}
\label{fig:2d_resolution_errors}
\end{figure}

Next, we examine how the relative $L^2$ error depends on the collocation resolution. The top panel of Figure~\ref{fig:2d_resolution_errors} shows the relative $L^2$ error as a function of training resolution. Increasing the collocation resolution from 80 to 380 points reduces the relative $L^2$ error from order $10^{-4}$ to order $10^{-8}$. As in the one-dimensional problem, this reduction is substantial but non-monotonic. A similar trend is observed for different random initialization seeds (Appendix~\ref{sec:appendix_seed_dependence}).

The bottom panel of Figure~\ref{fig:2d_resolution_errors} shows the pointwise relative errors along the $z$ axis, with the two dotted vertical lines marking the puncture positions. The largest errors occur near the punctures, decreasing from order $10^{-2}$ at resolution $98$ to order $10^{-6}$ at resolution $380$. Away from these localized peaks, the higher-resolution profiles exhibit comparable error levels, typically of order $10^{-7}$. However, this similarity does not imply that further refinement ceases to improve the overall solution. Although the $z$-axis profiles do not exhibit the systematic convergence typically observed in finite-difference solvers, the relative $L^2$ errors continue to decrease with increasing resolution.

\subsection{Three-dimensional binary-puncture initial data}
\label{sec:three_dimensional_results}

We now consider the GW150914-like binary used to validate \nrpyell in Ref.~\cite{assumpcao_fast_2022}. The punctures are located at $z=\pm5M$, with individual ADM masses satisfying $M_{+}/M_{-}=36/29$, where $M=M_{+}+M_{-}$. The spins are $\mathbf{S}_{+}=0.31M_{+}^2\hat{\mathbf{y}}$ and $\mathbf{S}_{-}=-0.46M_{-}^2\hat{\mathbf{y}}$ for the positive- and negative-$z$ punctures, respectively.\footnote{The corresponding bare masses are $(m_{+},m_{-})\simeq(0.518420,0.391936)M$. The quasicircular orbital momenta are specified by $\mathbf{P}_{+}=-\mathbf{P}_{-}=(P_\phi,0,P_r)$, with $P_\phi\simeq9.53\times10^{-2}M$ and $P_r\simeq-8.45\times10^{-4}M$.} This configuration breaks axial symmetry, requiring a fully three-dimensional solution, $u=u(\xxzero,\xxone,\xxtwo)$. We set the focal parameter to $b=5M$, matching the puncture positions, and impose the first-order Robin condition at the outer boundary.

The fully three-dimensional problem presents a more demanding optimization task for two reasons. First, multiplication by $\sqrt{\hat{\gamma}}$ does not remove all coordinate-singular coefficients from the \SinhSymTP residual. Using the radial map $\tilde r(\xxzero)$ defined in Sec.~\ref{sec:coordinates-systems}, the rescaled residual is
\begin{equation}
\begin{aligned}
\hat{\mathcal R}[u] ={}&
\frac{\partial}{\partial\xxzero}\left[
\frac{\tilde r\sqrt{\tilde r^{\,2}+b^2}\sin\xxone}{\tilde r'}
\frac{\partial u}{\partial\xxzero}\right]
\\
&+\frac{\partial}{\partial\xxone}\left[
\frac{\tilde r\tilde r'\sin\xxone}{\sqrt{\tilde r^{\,2}+b^2}}
\frac{\partial u}{\partial\xxone}\right]
\\
&+\frac{\tilde r'(\tilde r^{\,2}+b^2\sin^2\xxone)}
{\tilde r\sqrt{\tilde r^{\,2}+b^2}\sin\xxone}
\frac{\partial^2 u}{\partial\xxtwo^2}
\\
&+\frac{\sqrt{\hat{\gamma}}\tilde A_{ij}\tilde A^{ij}}
{8(\psi_{\rm singular}+u)^7},
\end{aligned}
\label{eq:3d_rescaled_residual}
\end{equation}
where a prime denotes differentiation with respect to $\xxzero$. The azimuthal-derivative coefficient retains inverse powers of $\sin\xxone$ and $\tilde r$, unlike the radial and polar derivative coefficients. In the axisymmetric case, $\partial^2u/\partial\xxtwo^2=0$, so this term vanishes identically. These are coordinate singularities, not physical singularities of the solution. However, a randomly initialized network in native coordinates does not automatically satisfy the corresponding regularity conditions. The large coefficient can therefore amplify its azimuthal derivatives, making the initial residual more sensitive to initialization than in the axisymmetric case.

The second challenge is the memory required to evaluate the full collocation grid and its derivatives on a single GPU, the training mode currently supported by \pinndorama. The additional spatial dimension restricts the resolutions accessible within this memory limit. During our three-dimensional tests, we did not establish the clear decrease in error with increasing resolution found in the axisymmetric case.

\begin{figure}[h]
\centering
\includegraphics[width=\linewidth]{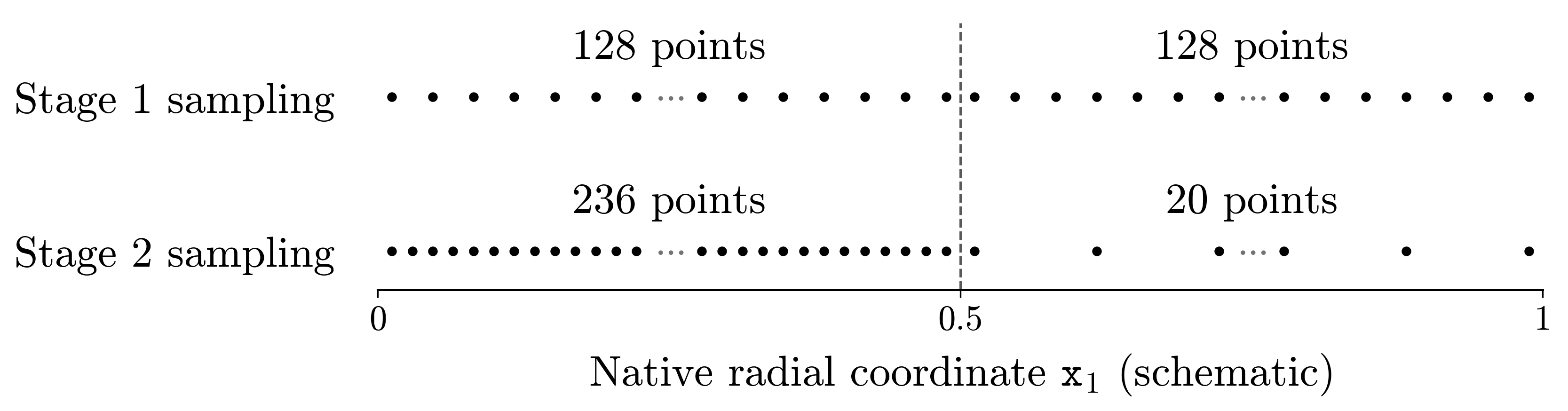}
\caption{Native radial sampling for SSBroyden stage 1 (uniform) and stage 2 (two-zone), with the same total number of points. The dashed line marks the zone boundary at $\xxzero=0.5$. Dots represent a subset of samples; displayed spacings are schematic.}
\label{fig:3d_sampling_distribution}
\end{figure}

To address these challenges, we divide SSBroyden training into two stages, retaining the network parameters while redistributing the same number of collocation points toward the inner region in stage 2. Figure~\ref{fig:3d_sampling_distribution} illustrates this strategy. The initial calculation uses a uniform cell-centered grid with $(\Nzero,\None,\Ntwo)=(256,256,16)$ and is trained with 10,000 Adam steps followed by 50,000 steps of SSBroyden stage 1.\footnote{We use a fixed seed of 503 to initialize the JAX pseudorandom key from which the initial network weights are drawn using He initialization; biases start at zero. The collocation grids are deterministic and do not depend on this seed. Stage 2 retains the trained parameters without reinitialization.}

\begin{figure}[h]
\centering
\includegraphics[width=\linewidth]{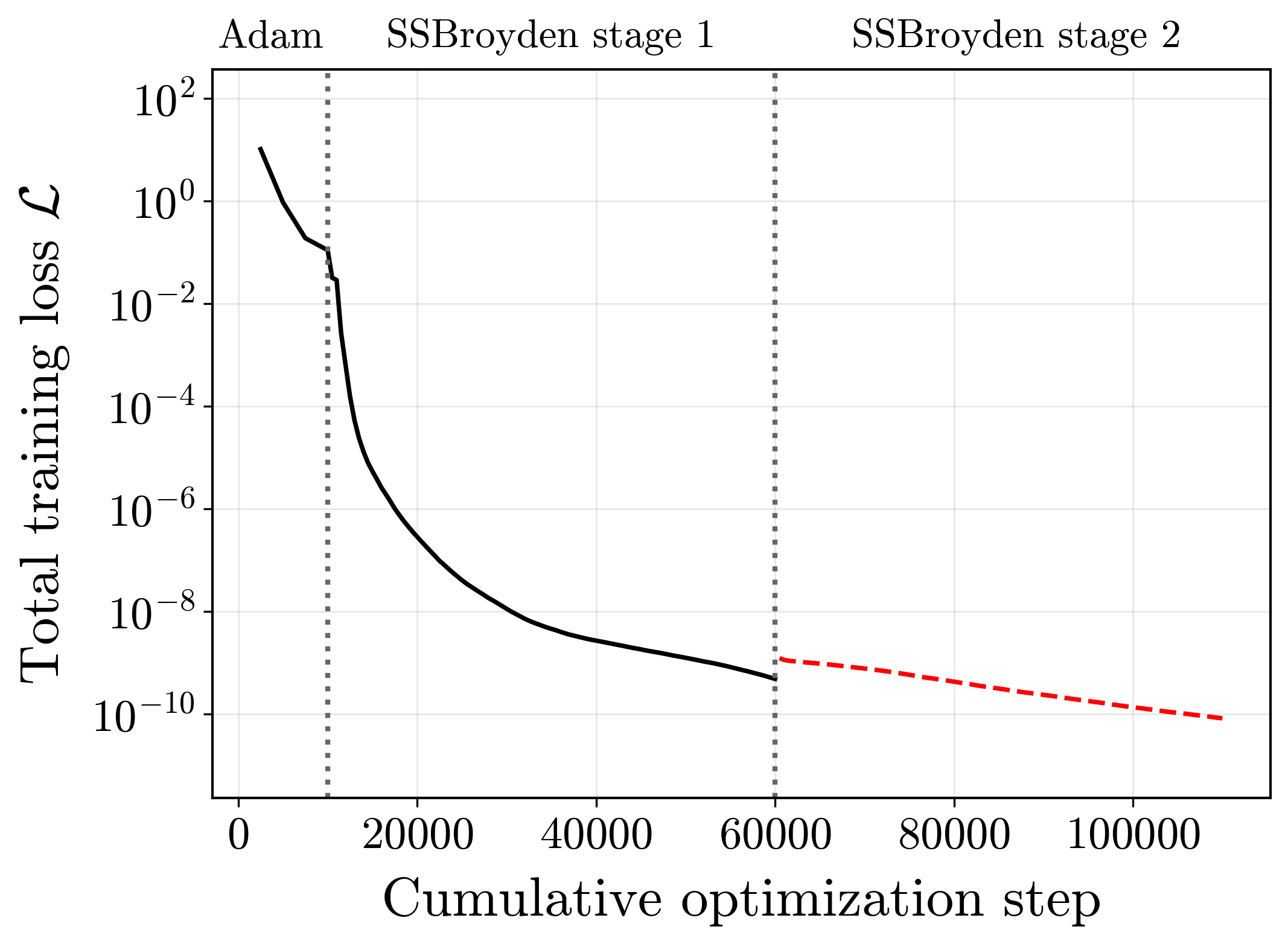}\par\smallskip
\includegraphics[width=\linewidth]{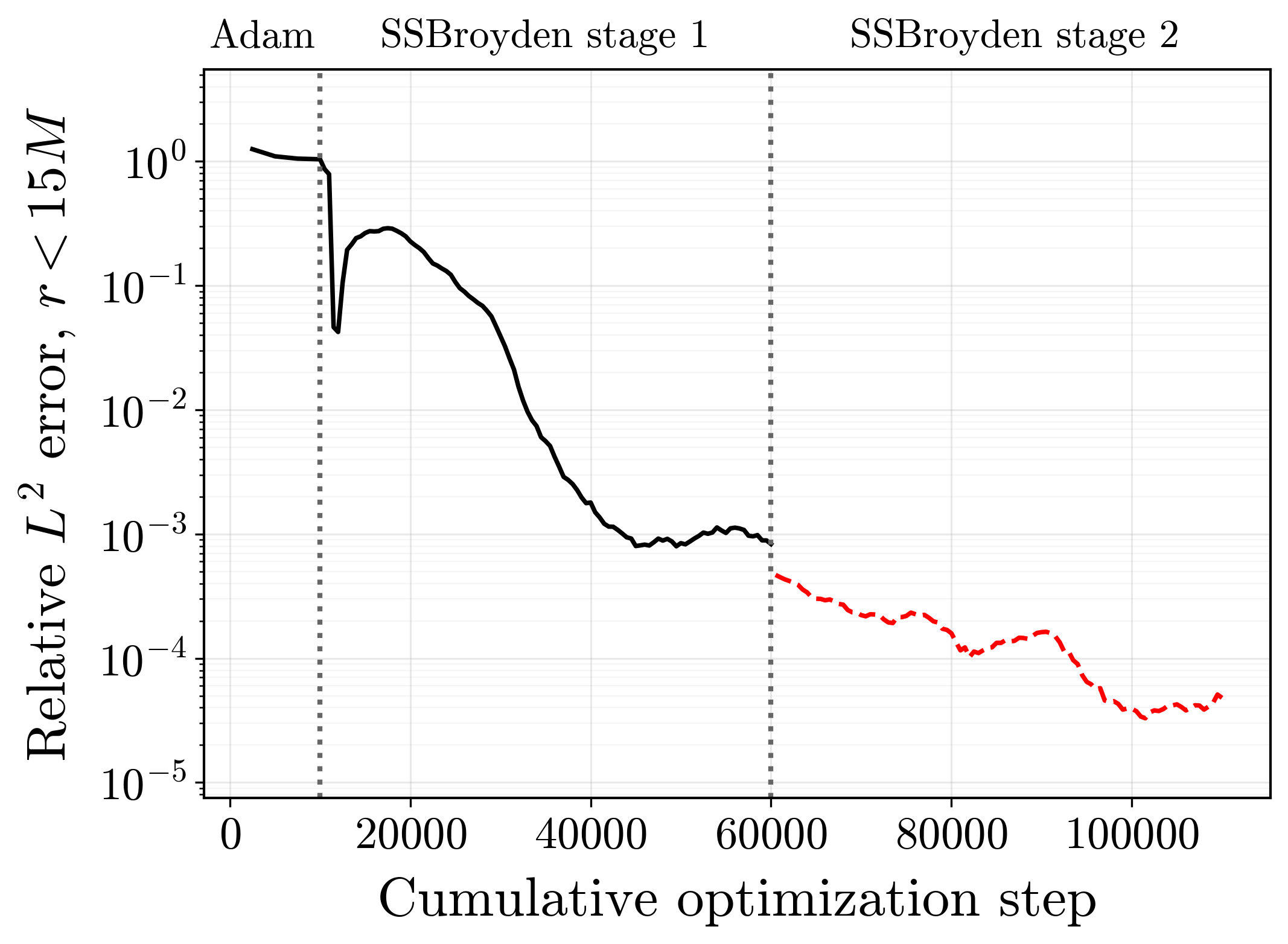}
\caption{Training loss (top) and relative $L^2$ error inside $r<15M$ (bottom) through Adam and SSBroyden stages 1 and 2. Solid black curves show Adam and SSBroyden stage 1 on the uniform grid; dashed red curves show SSBroyden stage 2 on the two-zone radial grid. The dotted vertical lines mark the start of SSBroyden stage 1 at step 10000 and stage 2 at step 60000.}
\label{fig:3d_training_history}
\end{figure}

In SSBroyden stage 2, we continue for a further 50,000 steps without repeating Adam. Only the native radial sampling changes: 236 cell centers are placed on $\xxzero\in[0,0.5]$ and 20 on $\xxzero\in[0.5,1]$, uniformly spaced within each interval. The inner half therefore receives 236 points instead of the original 128, while the angular sampling, network architecture, and total number of points remain unchanged. This allows us to increase the inner sampling density without increasing the size of the collocation grid.

To assess the improvement independently of the sampling redistribution, both stages are evaluated on the same $256\times256\times64$ native-coordinate grid inside $r<15M$. Doubling the evaluation resolution changes both final relative $L^2$ norms by less than $0.1\%$.

Figure~\ref{fig:3d_training_history} shows the training loss and relative $L^2$ error as functions of the cumulative number of optimization steps. As in the axisymmetric calculation, the relative error remains of order unity after Adam. SSBroyden stage 1 reduces it below $10^{-3}$, while stage 2 reaches a few times $10^{-5}$, improving the final accuracy by more than an order of magnitude. Redistributing the collocation points changes the discrete loss function, producing discontinuous changes in both the loss and relative error at the transition between stages 1 and 2. At this transition, the loss increases slightly, while the $L^2$ error decreases. Within each stage, the loss and relative error generally decrease together, although the relative error increases slightly near the end of stage 2 while the loss continues to decrease. Both SSBroyden stages reach the prescribed maximum of 50,000 optimization steps.

\begin{figure}[tbp]\centering\includegraphics[width=\linewidth]{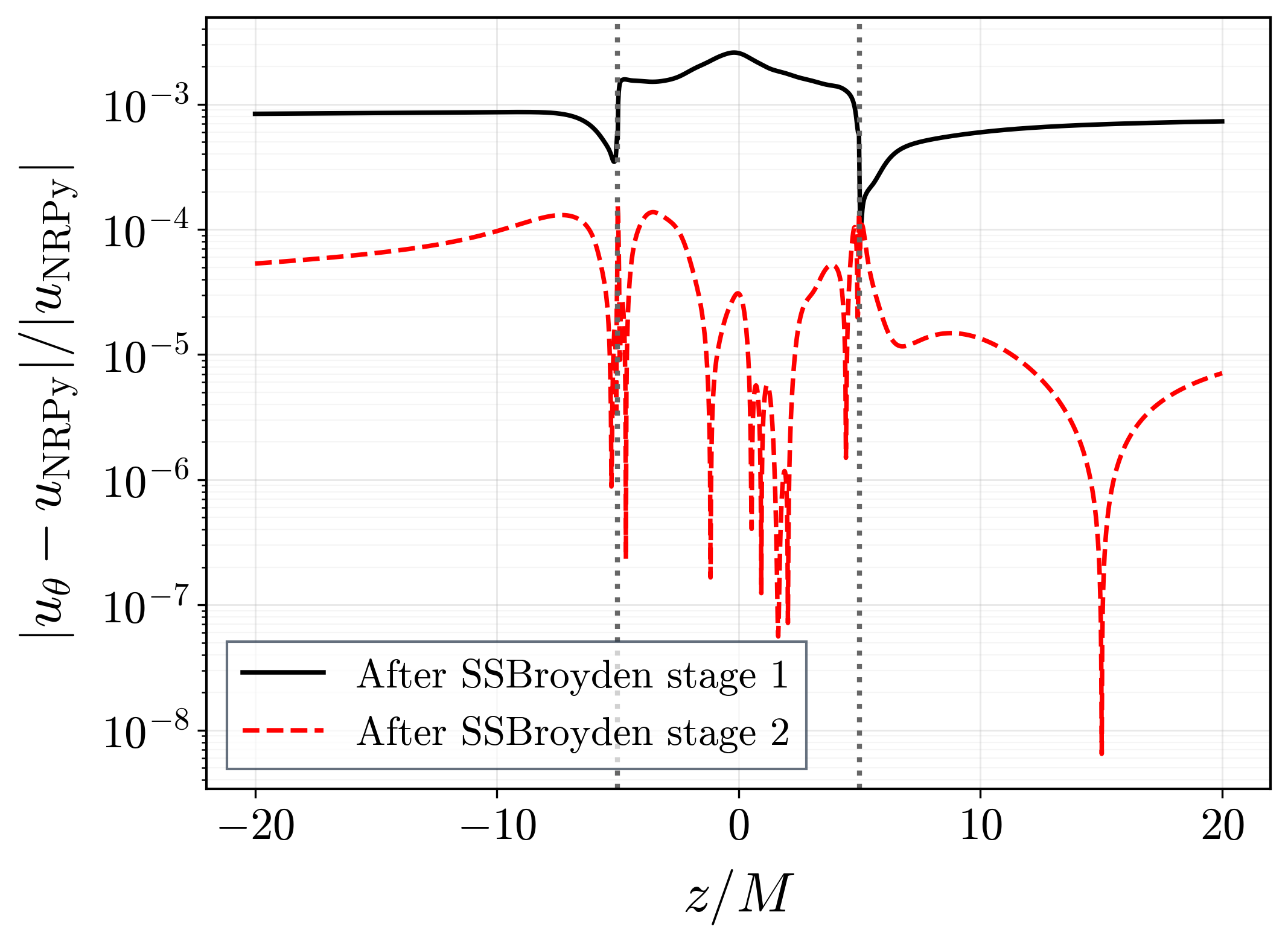}
\caption{Pointwise relative errors after SSBroyden stage 1 (solid black) and stage 2 (dashed red), evaluated over $-20\le z/M\le20$, with $x=y=10^{-6}M$. Vertical guides mark the punctures.}\label{fig:3d_z_axis_relative_error}\end{figure}

The axial profiles in Figure~\ref{fig:3d_z_axis_relative_error} show the corresponding spatial improvement. After SSBroyden stage 1, the largest sampled errors lie between the punctures and are of order $10^{-3}$. After stage 2, errors decrease over much of the displayed interval, with the largest remaining peak near the negative-$z$ puncture at order $10^{-4}$. Together with the volume norm, this demonstrates improved agreement with the numerical reference, although the accuracy remains lower than in the axisymmetric case.

\subsection{Parametric initial data and accelerated hyperbolic relaxation}
\label{sec:parametric_results}

Finally, we consider the neural-network solution as an initial guess for a numerical relaxation solver. We extend the axisymmetric model by including the common puncture spin as an additional input, $u_\theta=u_\theta(\xxzero,\xxone;S)$. The punctures have equal bare masses $m_+=m_-=M/2$, positions $z=\pm2.5M$, vanishing linear momenta, and equal axial spins $\mathbf{S}_+=\mathbf{S}_-=S\hat{\boldsymbol z}$. We train over $-0.2<S/M^2<0.2$, using a $128\times64$ spatial grid and $N_S=80$ uniformly spaced cell-centered spin samples. The network retains four hidden layers with forty neurons each. Training consists of 1,500 Adam steps followed by 50,000 SSBroyden steps, reaching a final loss of order $10^{-10}$.

\begin{figure}[tbp]
\centering
\includegraphics[width=\linewidth]{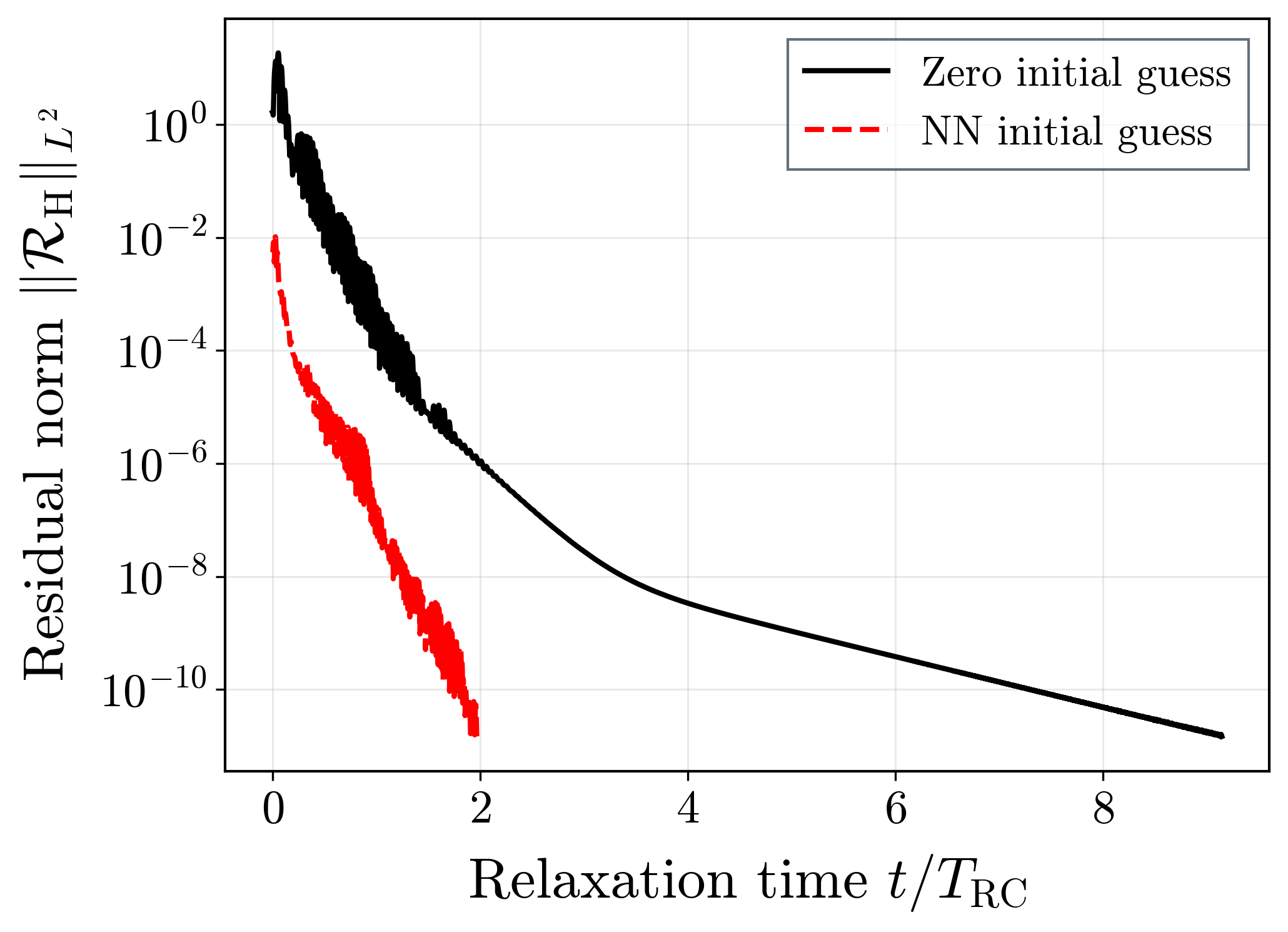}
\caption{Whole-domain Hamiltonian-residual $L^2$ norm during relaxation at $S=0.189M^2$, starting from $u=0$ (solid black) or the parametric neural-network prediction (dashed red). Time is normalized by the relaxation-wave-crossing time $T_{\rm RC}$. Both calculations use the same stopping criterion, based on the residual RMS inside $r<80M$, rather than the plotted whole-domain norm.}
\label{fig:parametric_relaxation_acceleration}
\end{figure}

To test this initial guess, we embed the trained network into the \nrpyell C executable and evaluate it at $S=0.189M^2$, a spin not included in the training samples. We compare relaxation from this prediction with relaxation from $u=0$, keeping the numerical grid, solver parameters, and time step unchanged. Figure~\ref{fig:parametric_relaxation_acceleration} shows the whole-domain $L^2$ norm of the Hamiltonian residual computed within the \nrpyell diagnostics. The horizontal axis is normalized by the relaxation-wave-crossing time $T_{\rm RC}$~\footnote{As defined in Appendix B of Ref.~\cite{assumpcao_fast_2022}.}, which is approximately $0.08864$ in code units for this configuration.

The neural-network initial guess reduces the initial residual norm from order unity to a few times $10^{-3}$, a reduction by a factor of approximately 300. Both calculations terminate when the residual RMS inside $r<80M$ falls below $10^{-14}$. Starting from zero requires approximately $9.1\,T_{\rm RC}$, whereas the network prediction reaches the same stopping criterion after approximately $2.0\,T_{\rm RC}$, with both whole-domain residual norms reaching order $10^{-11}$. The ratio of the measured relaxation times is $4.65$. This experiment suggests a promising hybrid approach in which parametric PINNs provide initial guesses that are subsequently refined by conventional numerical elliptic solvers.

\section{Conclusions and future work}
\label{sec:conclusions}
In this work, we presented \pinndorama, a physics-informed neural network solver for the Hamiltonian constraint in the puncture formalism. We tested the solver on problems of increasing complexity, beginning with a one-dimensional test problem with an exact solution. We then considered axisymmetric binary punctures with unequal spins and a fully three-dimensional GW150914-like binary. We validated the puncture solutions against \nrpyell. The calculations reach relative $L^2$ errors of order $10^{-6}$ in one dimension, $10^{-8}$ in axisymmetry, and $10^{-5}$ for the selected three-dimensional calculation. The networks are trained solely on the governing equation and boundary conditions, without an analytical guiding solution, which complements the approach of Ref.~\cite{619s-p6md}. In addition, our publicly available implementation provides a basis for independent assessment and further development of PINN methods for puncture initial data and beyond.

To achieve low errors, we used a combination of numerical and algorithmic choices. First, we sample the training points using the native \SinhSymTP coordinates of \nrpy, concentrating collocation points near the punctures while placing the outer boundary deep in the asymptotic region. We employ residual rescaling to control the magnitudes entering the loss, while rescaling the output to incorporate the expected asymptotic decay. In addition, we add penalty terms to the loss to enforce agreement between field values at native-coordinate labels representing the same Cartesian point. In the one-dimensional test, imposing the second-order Robin condition throughout an outer radial region substantially improves far-field accuracy. Most importantly, refining the Adam solution with the quasi-Newton optimizer SSBroyden reduces the error by several orders of magnitude. In three dimensions, continued optimization with collocation points redistributed toward the inner region at a fixed grid size improves accuracy by more than an order of magnitude.

Beyond providing standalone solutions, PINNs can also accelerate conventional solvers. A parametric network that takes the puncture spin as input, evaluated at a spin value not included in training, reduces the initial Hamiltonian residual norm in \nrpyell by a factor of $\sim300$ and the subsequent hyperbolic-relaxation time by a factor of $\sim4.6$. In this hybrid approach, the network supplies a physics-informed initial guess, while the numerical solver performs the final refinement to the desired tolerance. The reduction in relaxation time excludes network training; an overall computational benefit requires subsequent uses to offset this one-time cost. The same strategy could be tested with other iterative elliptic solvers.

Some limitations remain. The three-dimensional accuracy is lower than in the axisymmetric case, and we did not observe clear convergence with increasing resolution in 3D. Coordinate-singular coefficients of the azimuthal term and single-GPU memory constraints on the collocation resolution may contribute to these difficulties. In the 1D and 2D problems, the convergence with resolution is substantial but non-monotonic, and the results depend on the initialization seed. The displayed 1D and 2D training histories also suggest that the stopping criterion can be overly conservative, allowing optimization to continue after the relative error has plateaued. Future work will investigate multi-GPU training, adaptive sampling, stopping criteria, and regularity-aware representations and residual weightings that reduce coordinate-related scale disparities. We also plan to extend the parametric network to the full binary parameter space, including masses, momenta, spins, and separation, and to three dimensions, providing fast initial data or initial guesses for numerical-relativity simulations.

\section{Acknowledgments}
We acknowledge helpful discussions with Dimitra Tseneklidou, Yue Wu, and Steven Brandt. T.A. gratefully acknowledges support from NSF grants OAC-2229652 and AST-2108269 and from UKRI grant UKRI2492.
I.P. was supported by the London Mathematical Society (LMS), under the Undergraduate Research Bursary Scheme URB-2025-21. MB acknowledges partial support from the STFC Consolidated Grant nos. ST/Z000424/1, UKRI2492 and ST/C004235/1.
We acknowledge the EuroHPC Joint Undertaking for awarding this project access to the EuroHPC supercomputers LUMI-C, hosted by CSC (Finland) and the LUMI consortium, and MareNostrum 5, hosted by BSC (Spain), through a EuroHPC Extreme Scale Access call, under project EHPC-EXT-2025E02-070. This work used the DiRAC Extreme Scaling service (Tursa) at the University of Edinburgh and Memory Intensive service Cosma8 at Durham University, managed by the EPCC on behalf of the \href{https://www.dirac.ac.uk}{STFC DiRAC HPC Facility}. The DiRAC service at Edinburgh was funded by BEIS, UKRI and STFC capital funding and STFC operations grants. DiRAC is part of the UKRI Digital Research Infrastructure.
This work was also supported in part by a grant of access to OpenAI models through the ChatGPT for Academic Researchers program.

\section*{Code availability}
The \pinndorama source code, example configurations, and evaluation tools are available at \url{https://github.com/assumpcaothiago/Pinndorama}.

\appendix

\section{Derivation of the second-order Robin boundary condition}
\label{sec:appendix_second_order_robin}

The main text shows that extending the Robin condition through order
$r^{-2}$ requires the coefficient $D$. We retain $m$ explicitly below;
$m=1$ for the test problem considered here. To determine $D$, we insert the
large-radius expansion
\begin{equation}
    u(r)=\frac{C}{r}+\frac{D}{r^2}
    +\mathcal{O}(r^{-3})\, ,
    \label{eq:1d_asymptotic_expansion}
\end{equation}
where $C$ and $D$ are constants.  Its spherical Laplacian is
\begin{equation}
    \partial^2_r u+\frac{2}{r}\partial_r u
    =\frac{2D}{r^4}+\mathcal{O}(r^{-5})\, .
    \label{eq:1d_asymptotic_laplacian}
\end{equation}
For the source specified for the 1D problem, the large-radius expansions
\begin{equation}
    A(r)=\frac{32m}{r^4}+\mathcal{O}(r^{-5})\, ,
    \qquad
    \psi_{\rm singular}+u=1+\mathcal{O}(r^{-1})
\end{equation}
imply that the nonlinear term in
\eqref{eq:1d-toy-pde} satisfies
\begin{equation}
    \begin{aligned}
    N(r,u,m)&\equiv\frac{A(r)}{8[\psi_{\rm singular}(r)+u(r)]^7}\\
    &=\frac{4m}{r^4}+\mathcal{O}(r^{-5})\, .
    \end{aligned}
    \label{eq:1d_asymptotic_source}
\end{equation}
Substituting \eqref{eq:1d_asymptotic_laplacian}
and~\eqref{eq:1d_asymptotic_source} into
$\partial^2_r u+2\partial_r u/r+N=0$, the coefficient of $r^{-4}$ gives
\begin{equation}
    2D+4m=0\, ,
\end{equation}
so that $D=-2m$, while the leading coefficient $C$ remains undetermined by
the local asymptotic analysis.  The combination appearing in the second-order
condition then satisfies
\begin{equation}
    r^3\left(\partial_r u+\frac{u}{r}\right)
    =-D+\mathcal{O}(r^{-1})\, .
\end{equation}
Substituting $D=-2m$ yields
\begin{equation}
    r^3\left(\partial_r u+\frac{u}{r}\right)-2m
    =\mathcal{O}(r^{-1})\, ,
\end{equation}
which is the second-order residual in
\eqref{eq:second_order_robin}.  It is asymptotic rather than an exact
finite-radius identity.  Moreover, the coefficient $-2m$ is specific to the
1D test source, which is why this second-order operator is not imposed in the
2D or 3D puncture problems.

\section{Initialization-seed dependence}
\label{sec:appendix_seed_dependence}

\begin{figure}[h]
\centering
\includegraphics[width=\linewidth]{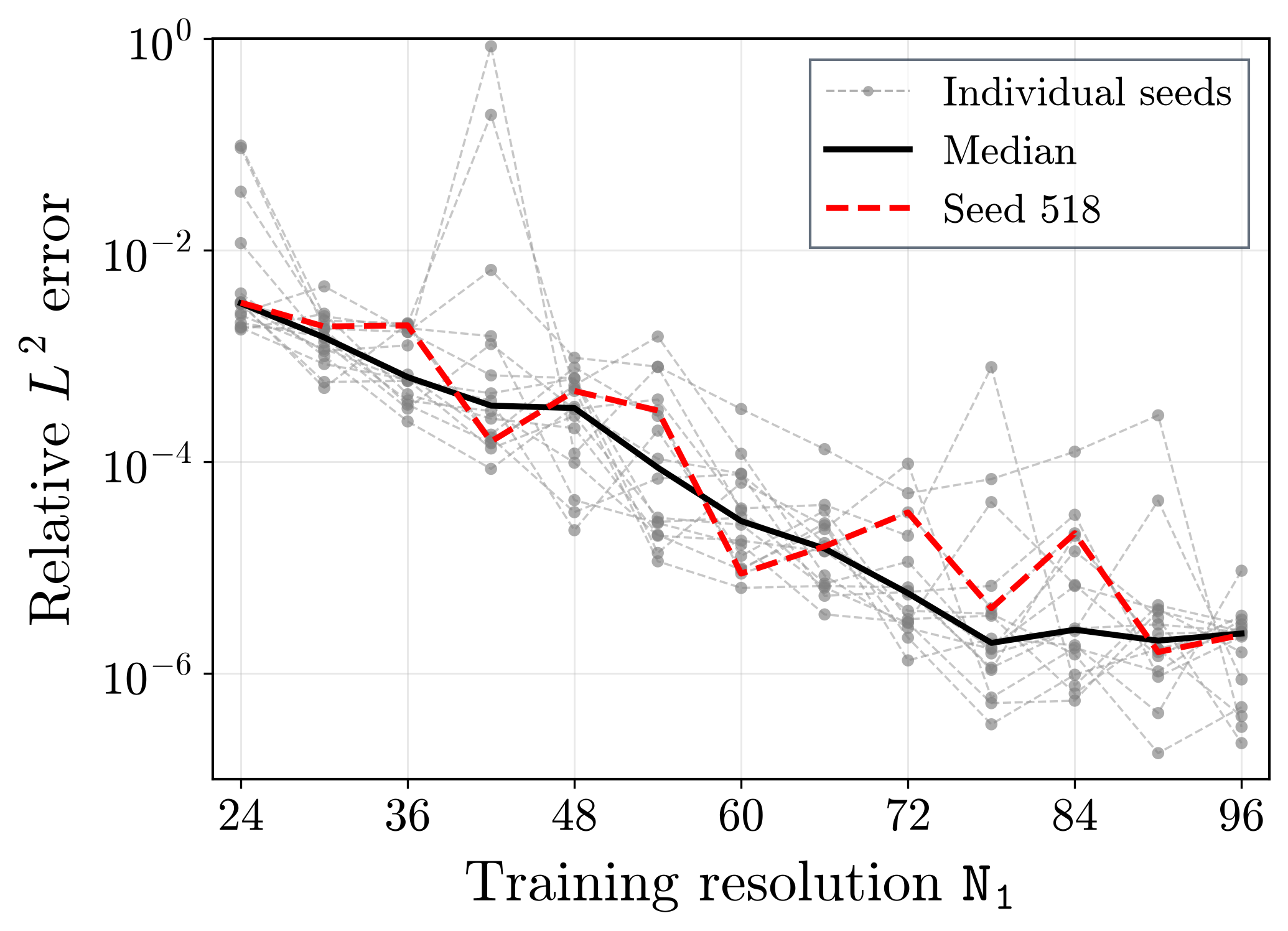}\par\smallskip
\includegraphics[width=\linewidth]{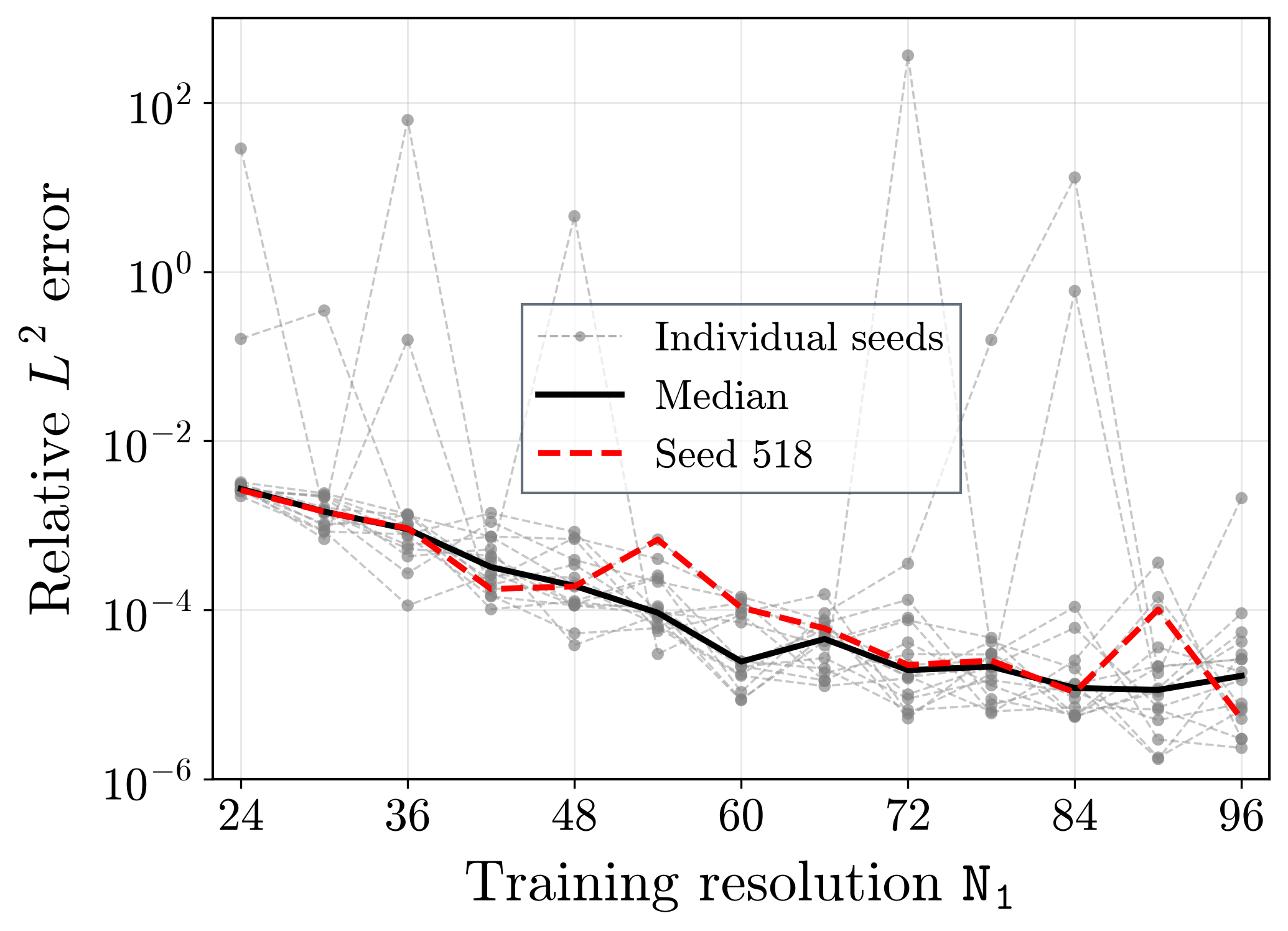}
\caption{Initialization-seed dependence of the one-dimensional relative $L^2$ error for the first-order Robin boundary condition (top) and the second-order Robin boundary condition (bottom). Gray points show all 16 seeds at each resolution, with gray dashed lines connecting the same seed across resolutions. The black solid curve is the median; the red dashed curve is seed 518, used in the main text. All errors use the same independent 400-point grid.}
\label{fig:1d_seed_dependence}
\end{figure}

To assess the sensitivity of the one-dimensional results to initialization, we train networks with 16 seeds, 503--518, at each resolution $\Nzero=24,30,\ldots,96$, for both Robin boundary conditions. These 416 calculations use the same architecture, initialization prescription, and optimizer settings as Sec.~\ref{sec:one_dimensional_results}. The relative $L^2$ errors in Figure~\ref{fig:1d_seed_dependence} use the same independent 400-point evaluation grid as the main text.

The median error decreases substantially with resolution, reaching approximately $10^{-6}$ and $10^{-5}$ at $\Nzero=96$ for the first- and second-order conditions, respectively. However, individual seed sequences are non-monotonic, and some runs retain much larger errors, particularly for the second-order condition. Increasing resolution therefore improves the typical accuracy without ensuring successful optimization for every initialization.

Separately, we repeated the second-order $\Nzero=96$ calculation 80 times with seed 518 and otherwise identical settings. All repetitions exactly reproduced the original run's final network parameters and predictions on the same independent 400-point grid, yielding a relative $L^2$ error of approximately $5\times10^{-6}$. This check distinguishes sensitivity to initialization from execution variability: unlike the seed scan, it holds the initialization fixed. It establishes repeatability for this configuration and the tested software and hardware environment, not for all problems or platforms.

\begin{figure}[htbp]
\centering
\includegraphics[width=\linewidth]{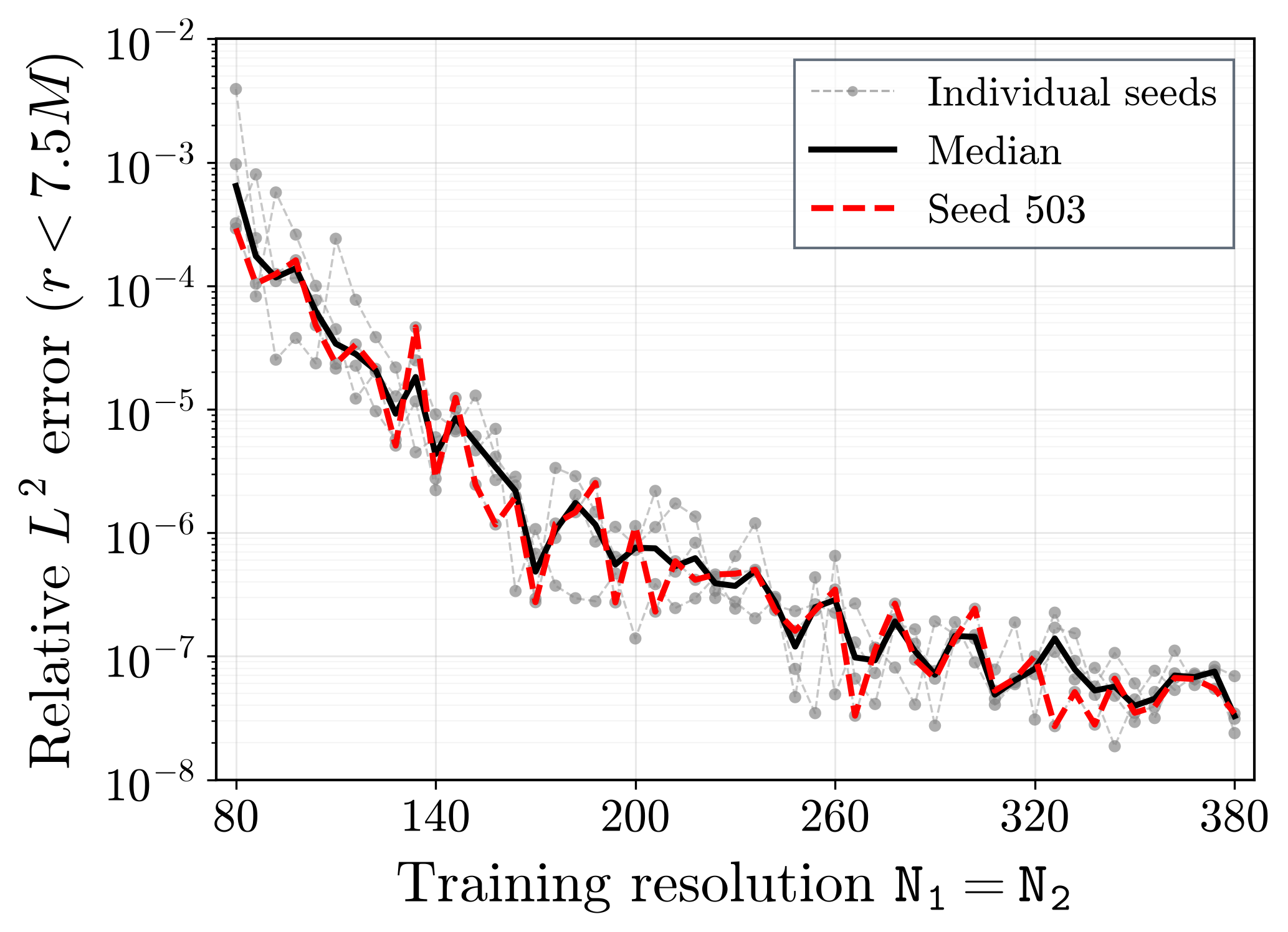}
\caption{Initialization-seed dependence of the axisymmetric relative $L^2$ error. Gray points show all four seeds, with gray dashed lines connecting the same seed across resolutions. The black solid curve is the median; the red dashed curve is seed 503, used in the main text. All errors use the same independent evaluation grid inside $r<7.5M$.}
\label{fig:2d_seed_dependence}
\end{figure}

For the axisymmetric problem, we compare four seeds, 503--506, at each of the 51 resolutions $\Nzero=\None=80,86,\ldots,380$, giving 204 calculations with the settings of Sec.~\ref{sec:axisymmetric_2d_results}. Figure~\ref{fig:2d_seed_dependence} uses the same independent $760\times760$ evaluation grid and region $r<7.5M$ as the main text.

The axisymmetric calculations exhibit a clearer resolution-convergence trend than the one-dimensional scans: the median error falls by approximately four orders of magnitude over the tested range, reaching order $10^{-8}$ at $\Nzero=\None=380$.

The highlighted curves used in the main text represent individual initializations rather than ensemble averages. Seed sensitivity remains a concern for PINNs~\cite{Lenau_2026} and neural networks~\cite{pmlr-v139-summers21a} in general. Its dependence on initialization, collocation sampling, and optimizer settings requires further investigation, and improving robustness to seed choice remains a goal for subsequent work.

\bibliography{references}

\end{document}